\documentclass[%
  reprint,
superscriptaddress,
showpacs,
 amsmath,
 amssymb,
 amsfonts,
 aps,
 pra,
showkeys
]{revtex4-1}

\RequirePackage{atbegshi} 

\usepackage{graphicx}       
\usepackage{dcolumn}        
\usepackage{bbold}          
\usepackage{verbatim}       
\usepackage[dvipsnames, table]{xcolor}
\usepackage{xparse}
\usepackage{siunitx}
\usepackage{subfigure}		
\renewcommand{\thesubfigure}{\alph{subfigure}}
\makeatletter
  \renewcommand{\@thesubfigure}{(\thesubfigure)\space}
  \def\@currentlabel{\p@subfigure\thesubfigure}
\makeatother

\usepackage{units}          
\usepackage[]{todonotes}    
\usepackage[matha]{mathabx} 
\usepackage{tabularx}
\usepackage{multirow}
\usepackage{booktabs}       

\usepackage[utf8]{inputenc}
\usepackage[T1]{fontenc}
\usepackage[ngerman,english]{babel}
\usepackage{slashed}
\usepackage{upgreek}
\usepackage{dsfont}

\usepackage{amsmath}
\usepackage{url}

\usepackage{hyperref}
\hypersetup{
    colorlinks=true,
    citecolor=black,
    linkcolor=black,
    urlcolor=black,
    anchorcolor=black,
    linktocpage
}

\DeclareMathOperator{\sech}{sech}

\newcommand\underrel[3][]{\mathrel{\mathop{#3}\limits_{%
      \ifx c#1\relax\mathclap{#2}\else#2\fi}}}

\newcommand{\NSigma}{N_\sigma}
\newcommand{\NTau}{N_\tau}

\newcommand{\Nf}{N_\text{f}}
\newcommand{\IntNf}{\Nf^\text{I}}

\newcommand{\Loewe}{LOEWE-CSC}
\newcommand{\Lcsc}{L-CSC}

\newcommand{\clqcd}{CL\kern-.25em\textsuperscript{2}QCD}
\newcommand{\Ocl}{OpenCL}
\newcommand{\Amd}{AMD }

\newcommand{\psibar}{\bar{\psi}}
\newcommand{\chiralcond}{\langle \psibar \psi \rangle}
\newcommand{\mpi}{m_{\pi}}
\newcommand{\ms}{m_{s}}
\newcommand{\mud}{m_{u,d}}
\newcommand{\Action}{\mathcal S}
\newcommand{\SGluon}{\Action_{\text{G}}}

\newcommand{\Binder}{B_4}
\newcommand{\Skewness}{B_3}

\newcommand{\ZTwoUniversality}{Z_2}
\newcommand{\NfTricr}{\Nf^\text{tric}}

\newcommand{\Link}{U}

\newcommand{\LatMassStaggered}{m}
\newcommand{\LatMassStaggeredZTwo}{\LatMassStaggered_{\ZTwoUniversality}}

\begin{document}

\title{The QCD chiral phase transition from non-integer numbers of flavors}


\author{Francesca Cuteri}
\email{cuteri@th.physik.uni-frankfurt.de}
\affiliation{
 Institut f\"{u}r Theoretische Physik, Goethe-Universit\"{a}t Frankfurt\\
 Max-von-Laue-Str.\ 1, 60438 Frankfurt am Main, Germany
}

\author{Owe Philipsen}
\email{philipsen@th.physik.uni-frankfurt.de}
\affiliation{
 Institut f\"{u}r Theoretische Physik, Goethe-Universit\"{a}t Frankfurt\\
 Max-von-Laue-Str.\ 1, 60438 Frankfurt am Main, Germany
}
\affiliation{
John von Neumann Institute for Computing (NIC)
GSI, Planckstr.\ 1, 64291 Darmstadt, Germany
}

\author{Alessandro Sciarra}
\email{sciarra@th.physik.uni-frankfurt.de}
\affiliation{
 Institut f\"{u}r Theoretische Physik, Goethe-Universit\"{a}t Frankfurt\\
 Max-von-Laue-Str.\ 1, 60438 Frankfurt am Main, Germany
}

\date{June 20, 2018}

\begin{abstract}
    Attempts to extract the order of the chiral transition of QCD at zero chemical potential, with two dynamical flavors of massless quarks, from simulations with progressively decreasing pion mass have remained inconclusive because of their increasing numerical cost.
    In an alternative approach to this problem, we consider the path integral as a function of continuous number $\Nf$ of degenerate quarks.
    If the transition in the chiral limit is first-order for $\Nf\ge3$, a second-order transition for $\Nf=2$ then requires a tricritical point in between.
    This in turn implies tricritical scaling of the critical boundary line between the first-order and crossover regions as the chiral limit is approached.
    Non-integer numbers of fermion flavors are easily implemented within the staggered fermion discretization.
    Exploratory simulations at $\mu=0$ and $\Nf = 2.8, 2.6, 2.4, 2.2, 2.1$, on coarse $\NTau = 4$ lattices, indeed show a smooth variation of the critical mass mapping out a critical line in the $(m,\Nf)$-plane.
    For the smallest masses the line appears consistent with tricritical scaling, allowing for an extrapolation to the chiral limit.
\end{abstract}

\maketitle

\section{Introduction}

Knowledge of the nature of the chiral phase transition of QCD with two flavors of massless quarks is of great importance for further progress in various directions of particle and heavy ion physics.
Besides its theoretical relevance for the interplay of chiral and confining dynamics, it has also implications for the thermal transition of physical QCD.
The light quark mass values realized in nature are close to the chiral limit and the thermal transition might well be affected by remnants of the chiral universality class.
Furthermore, the chiral limit of the two-flavor theory constrains the nature of the QCD phase diagram at finite baryon density, see e.g.~\cite{Rajagopal:2000wf}.

Unfortunately, simulations of lattice QCD with standard methods are impossible in the chiral limit because of the associated zero modes of the Dirac operator.
Thus, the computational cost rises with some power of the inverse quark mass and simulations are limited by some smallest mass depending on the lattice action, algorithms and machines.
Moreover, for the cheaper Wilson and staggered  type actions, the continuum limit should be taken before the chiral limit in order to avoid lattice artefacts.
As a result, the nature of the chiral transition in the continuum is not yet settled to date.

\begin{figure*}[t]
    \centering
    \subfigure[First order scenario in the $(\ms,\mud)$-plane]
        {\label{fig:firstOrderScenarioCP}\includegraphics[width=0.88\columnwidth,clip]{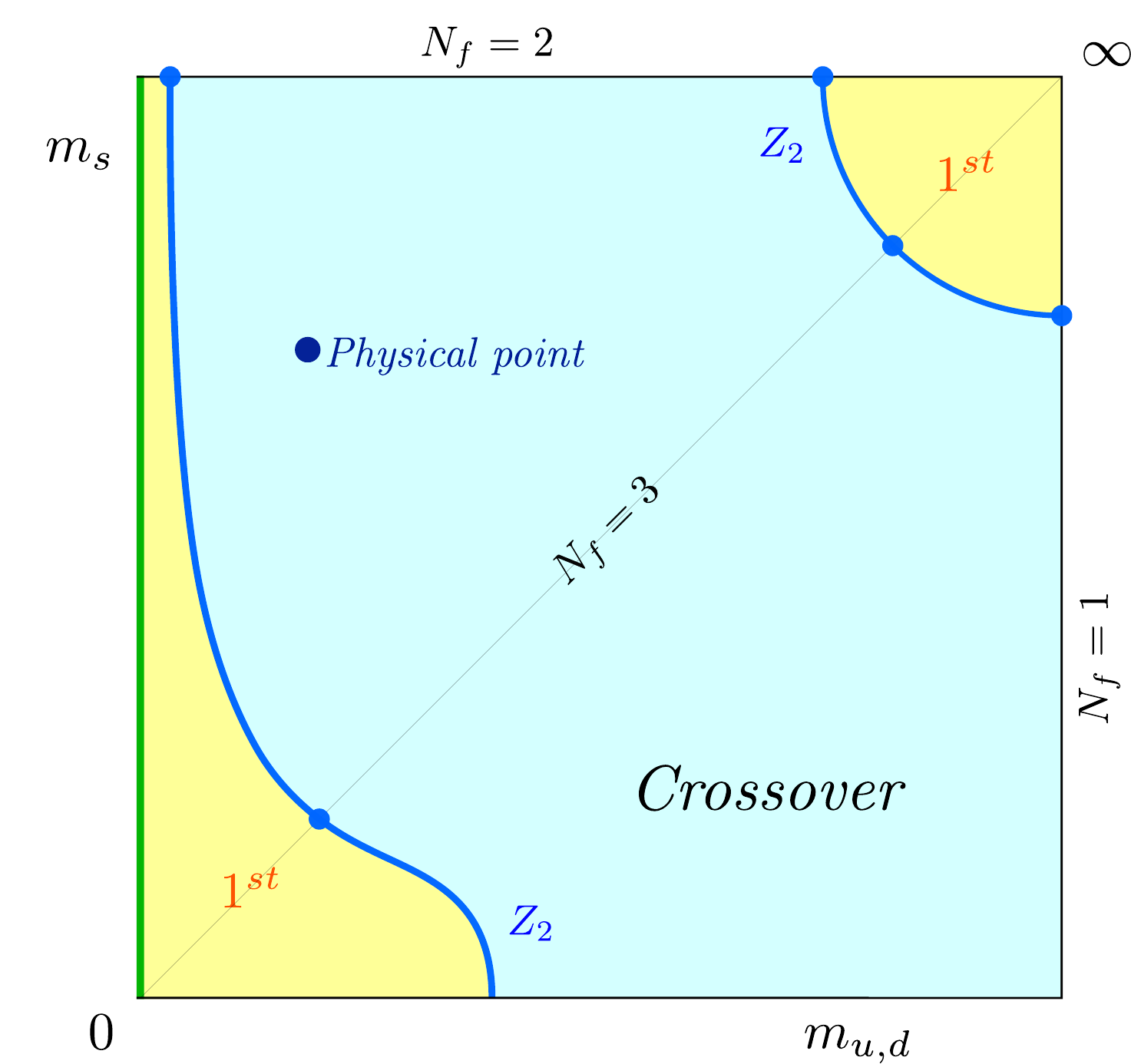}}
    \qquad
    \subfigure[Second order scenario in the $(\ms,\mud)$-plane]
        {\label{fig:secondOrderScenarioCP}\includegraphics[width=0.88\columnwidth,clip]{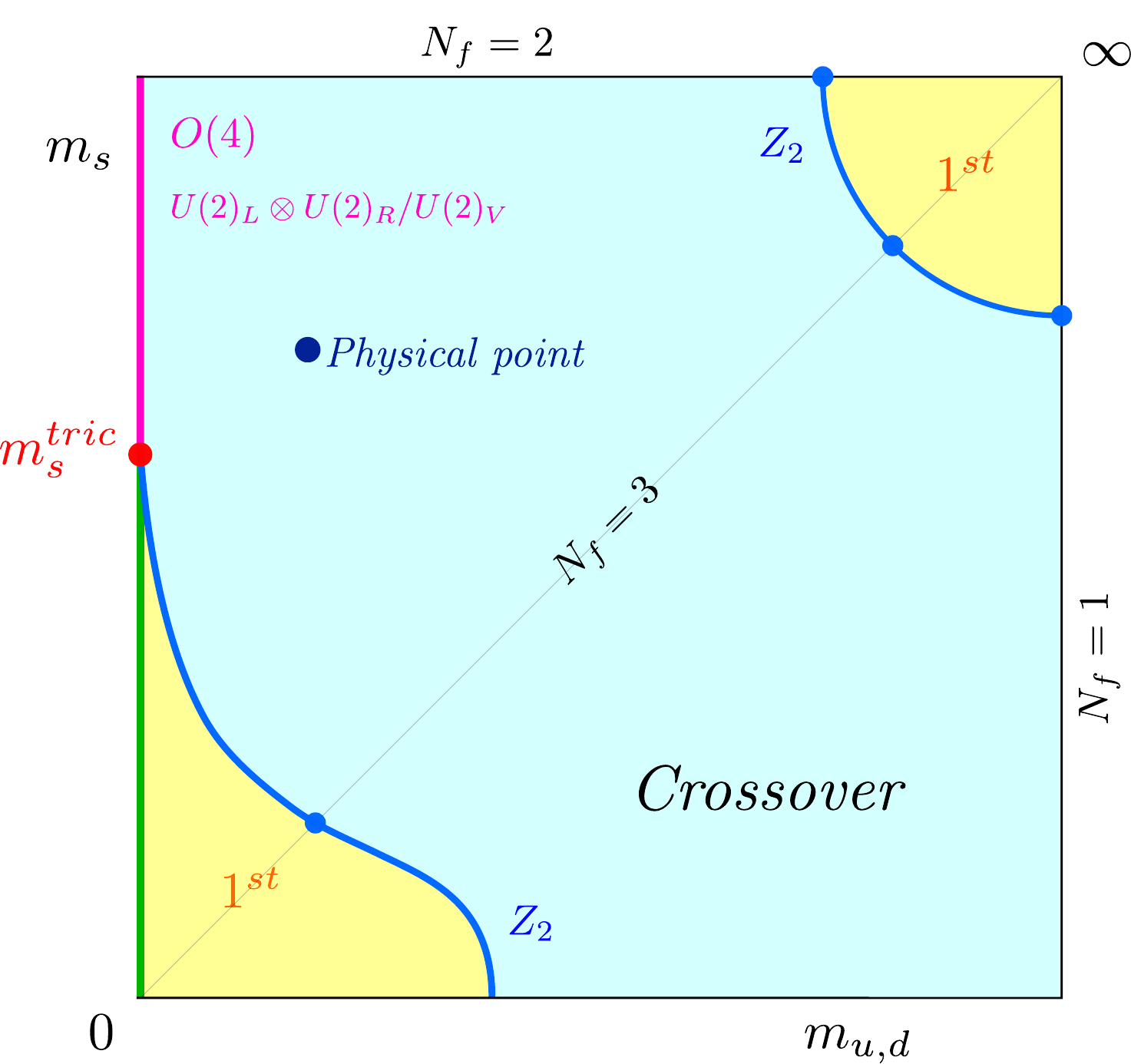}}
    \caption{Two possible scenarios for the order of the QCD thermal phase transition as a function of the quarks masses.\label{fig:scenariosCP}}
\end{figure*}

On the other hand, the nature of the thermal QCD transition is known to depend on the number of quark flavors $\Nf$ and on the masses $m$ of the quarks.
At zero baryon density and in the chiral ($m\rightarrow 0$) limit the transition is expected to be first-order for $\Nf=3$ degenerate quark flavors~\cite{Pisarski:1983ms}, while its nature for $\Nf=2$ degenerate quark flavors of mass $\mud$ is still under debate.
Based on an $\varepsilon$-expansion it was argued already long ago that the order of the transition depends crucially on the strength of the $U(1)_A$ anomaly at finite temperature~\cite{Pisarski:1983ms}.
Possible scenarios are depicted in Figure~\ref{fig:scenariosCP}.
Distinguishing between these for $\Nf=2$ in numerical simulations is a challenging task.
On coarse $\NTau=4$ lattices with lattice spacing $a\sim 0.3$ fm using unimproved Wilson~\cite{Philipsen:2016hkv} and staggered~\cite{Bonati:2014kpa} actions, a region of first-order transitions is explicitly visible, terminating in a critical quark mass in the $\ZTwoUniversality$-universality class.
For three degenerate flavors, $\Nf=3$, the first order region is wider than for $\Nf=2$, i.e.~the critical mass marking the boundary between the first-order and crossover regions is larger, both for Wilson~\cite{Jin:2015taa} and staggered fermions~\cite{Karsch:2001nf, deForcrand:2003vyj}.
This is in accord with the general expectation that a larger symmetry renders the transition stronger, as sketched in Figure~\ref{fig:scenariosMvsNf}.
Initial investigations of $\Nf=4$ in the staggered discretization confirm this general trend~\cite{deForcrand:2017cgb}.
On the other hand, again for both discretizations and all $\Nf$, the first order region shrinks as the lattice is made finer~\cite{deForcrand:2007rq,Jin:2017jjp}, with the critical quark masses becoming unreachably small when highly improved actions and/or fine lattices $\NTau\geq 12$ are used~\cite{Burger:2011zc, Brandt:2016daq, Ishikawa:2017nwl}.
Quite generally, for all $\NTau$'s investigated so far, the first-order region is vastly smaller for staggered discretizations, suggesting that the latter have the smaller cut-off effects in the critical quark mass configuration.
Altogether it thus remains a formidably difficult task to determine whether or not a finite first-order region survives in the continuum limit.

This situation motivates attempts to better constrain the first-order region by studying its extension in additional parameter directions, which might allow for controlled extrapolations to the chiral limit.
The idea is based on the fact that a first-order transition in the chiral limit on a finite system represents a 3-state coexistence (with the chiral condensate being positive, negative or zero).
If a continuous parameter is varied such as to weaken the transition, like increasing the strange quark mass $\ms$ in Figure~\ref{fig:secondOrderScenarioCP}, the 3-state coexistence may terminate in a tricritical point, which governs the functional behavior of the second-order boundary lines emanating from it by known critical exponents.
Thus, if such a boundary line can be followed into the tricritical scaling regime, an extrapolation becomes possible.
This kind of approach has been successfully tested varying imaginary chemical potential.
There is compelling numerical evidence~\cite{PhysRevD.83.054505,Pinke:2014cra,Cuteri:2015qkq,Philipsen:2016swy} that, whatever the realized scenario for the Columbia plot at $\mu=0$, the chiral first order region widens in $\mud$ as soon as a nonzero imaginary chemical potential is switched on, which makes it easier to map out its second order critical boundary.
Using unimproved staggered fermions on $\NTau=4$ lattices, tricritical scaling is observed and an extrapolation yields a first-order transition in the chiral limit~\cite{Bonati:2014kpa}.
A different analysis~\cite{Ejiri:2015vip} also finds the chiral first-order region to become larger both with increasing chemical potential and additional $\Nf$ heavy flavors considering $(2+\Nf)$-flavor QCD.

In this paper we investigate to which extent we can alternatively exploit the dependence of the chiral transition on the number of light degenerate flavors $\Nf$ as a means to perform controlled chiral extrapolations.
To this end, we treat $\Nf$ as a continuous real parameter as explained in Section~\ref{sec:nf} and sketched in Figure~\ref{fig:scenariosMvsNf}.
Starting from $\Nf=3$, where there is a first-order chiral transition region for finite quark masses, we then follow its boundary line to smaller $m$ and $\Nf$ until we indeed observe an apparent onset of tricritical scaling.
The extrapolation to the chiral limit with known exponents can then decide between the two scenarios in the figure, depending on whether the tricritical value of $\Nf$ is larger or smaller than 2.
The problem is analogous to that of determining the order of the phase transition of $q$-state Potts models in $d$ dimensions, which can be either first- or second-order separated by a tricritical line in $(q,d)$-space.
In particular, for $d=3$ the transition is first order for $q=3$ and second order for $q=2$.
In~\cite{Fortuin:1971dw} an analytic continuation to non-integer values of $q$ leading to simulable models was presented, for which a tricritical value $q_{tric}\approx 2.2$ was determined~\cite{Barkema:1991qx}.

In Section~\ref{sec:num} our numerical analysis using staggered fermions on $\NTau=4$ lattices is explained.
Section~\ref{sec:finiteSizeEffects} is dedicated to a study of the finite size effects affecting the skewness of the chiral condensate distribution sampled in our simulations.
In Section~\ref{sec:results} we present and discuss our results.
Lastly, Section~\ref{sec:conc} contains our conclusions.

\section{Lattice QCD for non-integer $\Nf$ \label{sec:nf}}

\begin{figure*}[t]
    \centering
    \subfigure[First order scenario in the $(\mud,\Nf)$-plane]
        {\label{fig:firstOrderScenarioMvsNf}\includegraphics[width=0.88\columnwidth,clip]{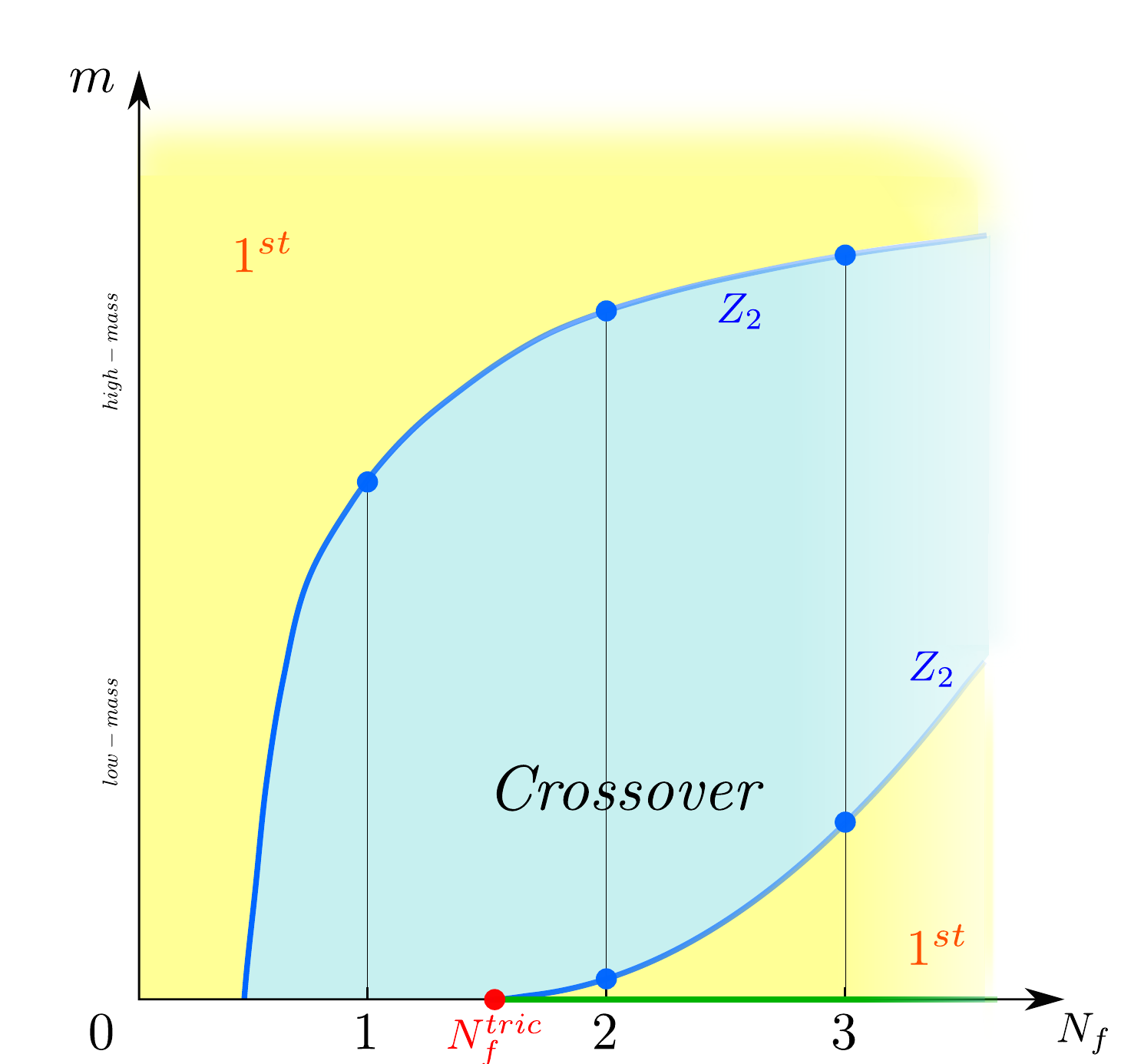}}
    \qquad
    \subfigure[Second order scenario in the $(\mud,\Nf)$-plane]
        {\label{fig:secondOrderScenariosMvsNf}\includegraphics[width=0.88\columnwidth,clip]{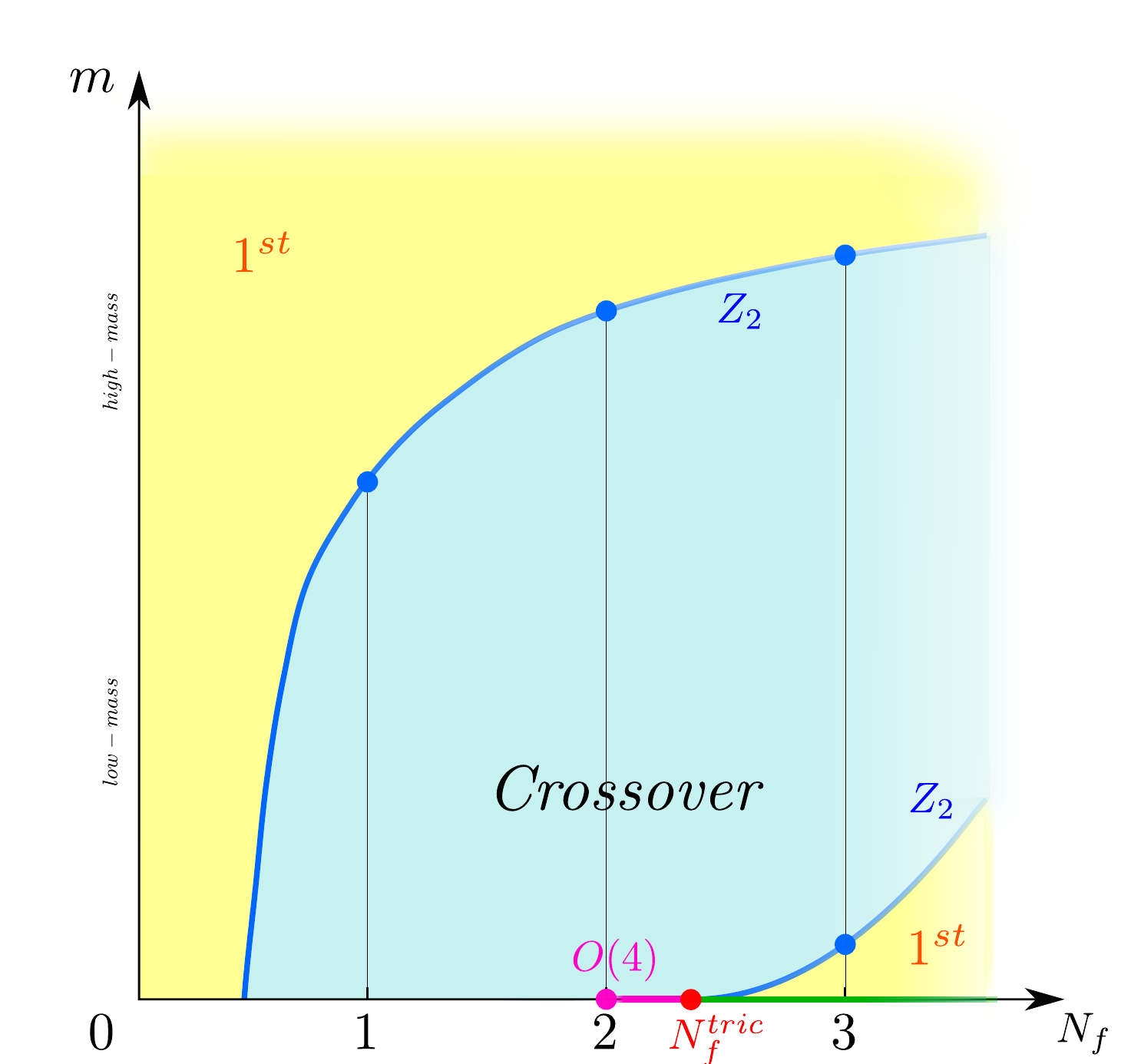}}
    \caption{The two considered possible scenarios for the order of the QCD thermal phase transition as a function of the light-quarks mass and the number of fermion flavors.\label{fig:scenariosMvsNf}}
\end{figure*}

We consider QCD with $\Nf$ mass-degenerate quarks of mass $m$ at zero density and the partition function reads
\begin{equation}
    Z_{\Nf}(m) = \int \mathcal{D}\Link \left[\det M(\Link,m)\right]^{\Nf} e^{-\SGluon}\;.
\end{equation}
Ignoring, for the moment, that it originates from QCD, we can formally view this as a partition function of some statistical system characterized by a continuous parameter $\Nf$.
Our question then is for which (tricritical) value of $\Nf$ the phase transition displayed by this system changes from first-order to second-order.

Of course, the extension of $Z_{\Nf}(m)$ to non-integer values of $\Nf$ is not unique, with infinitely many possibilities to \emph{fill in} values such that in the limits $\Nf\rightarrow 2,3$ the respective QCD partition functions are recovered.
In general such interpolations using non-integer powers of the determinant will not correspond to local quantum field theories.
One possibility, already suggested in~\cite{deForcrand:2015daa}, that does represent a local quantum field theory is to consider a partition function with a fixed integer number of $\IntNf+1$ flavors, $Z_{\IntNf+1}(m, m_1)$, where $\IntNf$ quarks are mass-degenerate with mass $m$ and an additional flavor has mass $m_1$ (the superscript in $\IntNf$ is there to stress that we consider an integer number of flavors in this case).
Starting from $m_1=m$, we can increase the mass $m_1$ until that flavor effectively decouples, such that the partition function for $\IntNf$ flavors is recovered.
For any value of $m_1$ in the local quantum field theory with fixed $\IntNf+1$, we can now find  an in general non-integer value of $\Nf$ for our statistical system $Z_{\Nf}(m)$, such that (with all other parameters held fixed)
\begin{equation}
    Z_{\IntNf+1}(m,m_1)=Z_{\Nf}(m) \;.
\end{equation}
For our practical exploration we employ a simpler strategy using staggered fermions, where the determinant needs to be rooted in order to describe any integer number of quark species smaller than four.
The RHMC algorithm~\cite{Kennedy:1998cu} is used to simulate any number $\Nf$ of degenerate flavors of staggered fermions, with $\frac{\Nf}{4}$ being the power to which the fermion determinant is raised in the lattice partition function.
Thus, even theories with integer $\IntNf=2,3$ suffer from an apparent lack of locality as long as no continuum limit is taken. 
In this framework it is straightforward to continuously vary $\Nf$, with exactly the same locality properties as for integer $\IntNf<4$ at any finite lattice spacing.

The precise value of $\NfTricr$, which we seek to determine, has no meaning in itself other than being located between two particular integer $\IntNf$'s.
In particular, for each lattice spacing the nature of the chiral phase transition for $\Nf=2$ can be decided by whether the extrapolated $\NfTricr$ is smaller or larger than 2, as in Figure~\ref{fig:scenariosMvsNf}. In the left scenario,
this produces a critical quark mass $\LatMassStaggeredZTwo(a)$ for the $\IntNf=2$ theory, which eventually can be continuum extrapolated.

We should emphasize that a formal proof for the locality of the target theories at non-integer $\Nf$ is missing, while properties connected with universality, such as the existence of second order boundaries and tricritical scaling, do depend on locality.
On the other hand, an effective locality, in the sense that all $n$-point functions decay exponentially, is sufficient for this purpose and, as we shall see, we do indeed observe the expected scaling phenomena, already on coarse lattices.

\section{Simulation setup and numerical strategy \label{sec:num}}

\newcolumntype{P}{>{\centering\arraybackslash}p{0.13\textwidth}}
\newcolumntype{Q}{>{\centering\arraybackslash}p{0.04\textwidth}}
\newcolumntype{R}{>{\centering\arraybackslash}p{0.08\textwidth}}
\renewcommand{\arraystretch}{1.2}
\begin{table*}[]
    \centering
    \begin{tabularx}{\textwidth}{QRP@{\qquad}*{5}{P}}
        \toprule[0.3mm]
        \multirow{2}{*}{$\Nf{}$} & \multirow{2}{*}{$m$} & \multirow{2}{*}{$\beta$ range} &
        \multicolumn{5}{c}{Total statistics per spatial lattice size $\NSigma{}$ $\bigl($ \texttt{\#} of simulated $\beta$ values)} \\
        & & & 8 & 12 & 16 & 20 & 24 \\
        \midrule[0.1mm]
        \multirow{4}{*}{2.8}
            & 0.0250 & 5.1620-5.1680 & 1.12M (4) & 0.84M (3) & 0.86M (3) &     -     &     -     \\
            & 0.0225 & 5.1580-5.1610 & 1.40M (4) & 0.80M (3) & 1.06M (3) &     -     &     -     \\
            & 0.0200 & 5.1520-5.1580 & 0.94M (4) & 1.16M (3) & 0.56M (3) &     -     &     -     \\
            & 0.0175 & 5.1490-5.1520 & 1.32M (4) & 1.20M (3) & 0.96M (3) &     -     &     -     \\
        \midrule[0.1mm] 
        \multirow{4}{*}{2.6} 
            & 0.0200 & 5.1840-5.1900 & 0.96M (4) & 0.92M (3) & 0.78M (3) &     -     &     -     \\
            & 0.0175 & 5.1790-5.1830 & 1.12M (4) & 1.17M (4) & 0.95M (3) &     -     &     -     \\
            & 0.0150 & 5.1740-5.1780 & 1.24M (3) & 1.12M (3) & 1.08M (3) &     -     &     -     \\
            & 0.0125 & 5.1680-5.1740 & 1.08M (4) & 0.92M (3) & 1.48M (4) &     -     &     -     \\
        \midrule[0.1mm] 
        \multirow{4}{*}{2.4}
            & 0.0150 & 5.2060-5.2120 & 1.04M (4) & 0.88M (3) & 0.68M (3) & 0.48M (4) &     -     \\
            & 0.0125 & 5.2010-5.2070 & 1.12M (4) & 0.96M (3) & 0.96M (3) & 0.32M (3) &     -     \\
            & 0.0100 & 5.1960-5.2020 & 0.92M (4) & 1.08M (3) & 1.04M (3) &     -     &     -     \\
            & 0.0075 & 5.1900-5.1960 & 1.40M (4) & 0.95M (3) & 1.03M (4) &     -     &     -     \\
        \midrule[0.1mm] 
        \multirow{4}{*}{2.2}
            & 0.0100 & 5.2280-5.2360 & 1.80M (4) & 2.00M (7) & 1.14M (4) & 0.96M (4) &     -     \\
            & 0.0075 & 5.2240-5.2300 & 1.24M (4) & 0.92M (3) & 1.12M (4) & 0.68M (3) &     -     \\
            & 0.0050 & 5.2200-5.2230 & 0.96M (4) & 0.84M (3) & 1.08M (4) & 1.04M (4) &     -     \\
            & 0.0025 & 5.2140-5.2180 & 1.16M (3) & 1.18M (5) & 0.76M (3) & 0.59M (3) &     -     \\
        \midrule[0.1mm] 
        \multirow{4}{*}{2.1}
            & 0.0045 & 5.2380-5.3400 &     -     & 0.68M (3) & 0.71M (3) & 0.74M (3) & 0.39M (3) \\
            & 0.0035 & 5.2340-5.2380 &     -     & 0.64M (3) & 0.82M (3) & 0.72M (3) & 0.30M (3) \\
            & 0.0025 & 5.2330-5.2360 &     -     & 1.00M (4) & 0.80M (3) & 0.66M (3) & 0.19M (3) \\
            & 0.0015 & 5.2310-5.2340 &     -     & 0.84M (4) & 0.74M (4) & 0.42M (4) &     -     \\
        \midrule[0.1mm] 
        \multirow{4}{*}{2.0}
            & 0.0020 & 5.2500-5.2520 &     -     & 0.67M (3) & 0.47M (3) & 0.16M (3) &     -     \\
            & 0.0015 & 5.2490-5.2510 &     -     & 0.72M (3) & 0.52M (3) & 0.28M (3) &     -     \\
            & 0.0010 & 5.2480-5.2500 &     -     & 0.68M (3) & 0.35M (3) & 0.07M (3) &     -     \\
            & 0.0007 & 5.2460-5.2490 &     -     & 0.82M (4) & 0.42M (3) & 0.08M (3) &     -     \\
        \bottomrule[0.3mm] 
  \end{tabularx}
  \caption{Overview of the statistics accumulated in all the simulations ($\NTau=4$ and $\mu=0$).
           The indicated $\beta$ ranges are the ones over which, for the smallest considered volume, simulated data for $\Skewness(\beta,m)$~and $\Binder(\beta,m)$ were used as input for the Ferrenberg-Swendsen reweighting~\cite{Ferrenberg:1989ui}, which interpolates between the simulated $\beta$ to allow for a more precise determination of $\beta_c$ and, thus, of the order of the transition.
           Per each $\beta$ range we also indicate the number of simulated $\beta$ because the resolution in $\beta$ changed both with $m$ and with $\NSigma$.
           In the indicated total statistics per $\NSigma$ we are summing up the number of simulated trajectories over 4 independent Markov chains and over all simulated $\beta$.
           The average length of each chain is then easily inferable.}
  \label{tab:simOverview}
\end{table*}

All numerical simulations have been performed using the publicly available~\cite{CL2QCD} \Ocl-based code \clqcd~\cite{Philipsen:2014mra}, which is optimized to run efficiently on \Amd  GPUs and provides, among others, an implementation of the RHMC algorithm for unimproved rooted staggered fermions.
In our exploratory study, conducted at zero chemical potential $\mu=0$, we kept the temporal extent of the lattices fixed at $\NTau = 4$.
The ranges in mass $m$ and gauge coupling constant $\beta$ of the investigated parameter space were dictated by our purpose of locating the chiral phase transition for values of the mass $m$ around the critical $\LatMassStaggeredZTwo$ value, with the temperature related to the coupling according to $T = 1/(a(\beta)\NTau)$.

Moreover, to locate and identify the order of the chiral phase transition a finite size scaling analysis of the third and fourth standardized moments of the distribution of the (approximate) order parameter is necessary.
The $\text{n}^{\text{th}}$ standardized moment for a generic observable $\mathcal{O}$ is expressed as
\begin{equation}
    B_n(\beta,m,\NSigma) = \frac{\left\langle\left(\mathcal{O} - \left\langle\mathcal{O}\right\rangle\right)^n\right\rangle}{\left\langle\left(\mathcal{O} - \left\langle\mathcal{O}\right\rangle\right)^2\right\rangle^{n/2}} \; .
\end{equation}
Being interested in the thermal phase transition in the chiral limit, we measured the chiral condensate $\chiralcond$ as an 
approximate order parameter.
In order to extract the order of the transition as a function of the quark mass and number of flavors, we considered the kurtosis $\Binder(\beta,m)$~\cite{Binder:1981sa} of the sampled $\chiralcond$ distribution, evaluated at the coupling $\beta_c$ for which we have vanishing skewness $\Skewness(\beta=\beta_c,m,\NSigma)=0$, i.e.~on the phase boundary.
In the thermodynamic limit $\NSigma \rightarrow \infty$, the kurtosis $\Binder(\beta_c,m,\NSigma)$ takes the values of 1 for a first order transition and 3 for an analytic crossover, respectively, with a discontinuity when passing from a first order region to a crossover region via a second order point. For the $3D$ Ising universality class of interest here, it takes the value $1.604$~\cite{Pelissetto:2000ek},
\begin{equation}
    \lim_{\NSigma \to \infty}\Binder(\beta_c,m,\NSigma) =
    \begin{cases}
        1, & 1^{st} \mbox{ order}\\
        1.604, & 2^{nd} \mbox{ order } \ZTwoUniversality\\
        3, & \mbox{crossover}
    \end{cases}.
\end{equation}
\begin{figure*}
    \centering
        {\label{fig:analyticDistributions}\includegraphics[width=.42\textwidth]{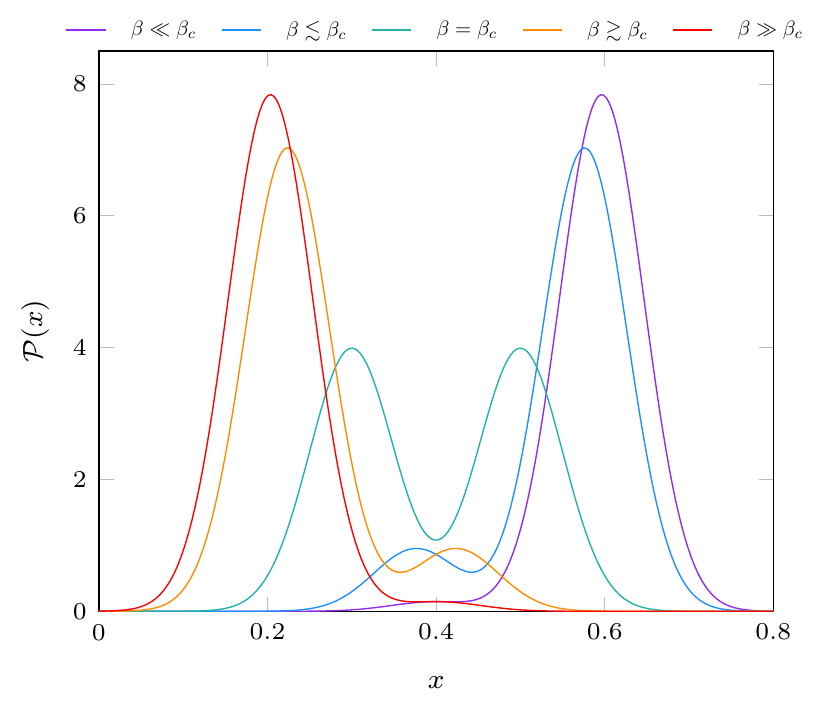}} \qquad
        {\label{fig:analyticSkewness}\includegraphics[width=.435\textwidth]{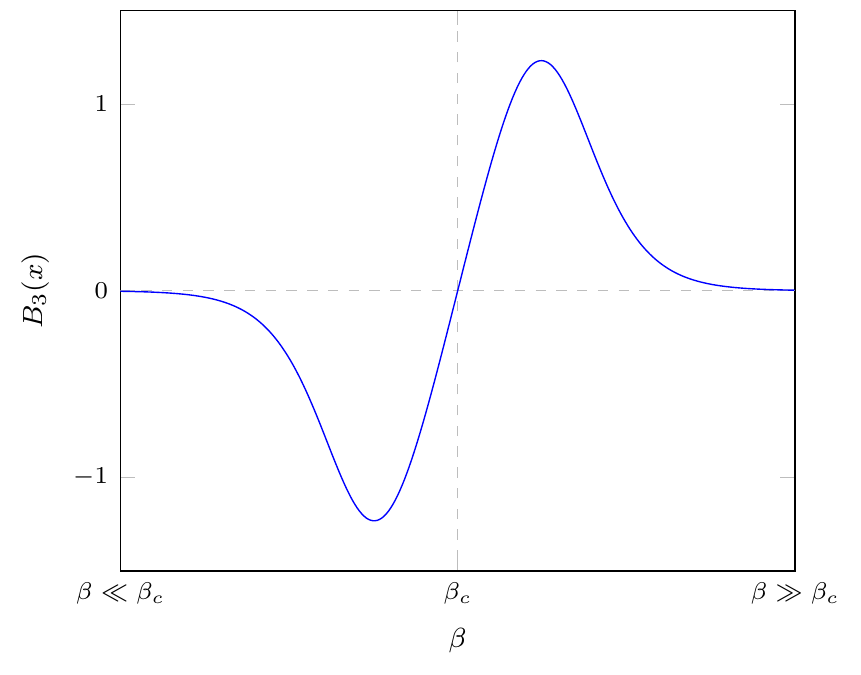}}
    \caption{$\mathcal{P}(x)$ at various $\beta$ values and the corresponding skewness as a function of $\beta$.}
    \label{fig:AnalyticModel}
\end{figure*}%
The discontinuous step function is smeared out to a smooth function as soon as a finite volume is considered.
In the vicinity of a critical point, the kurtosis can be expanded in powers of the scaling variable $x\equiv(m - \LatMassStaggeredZTwo) \NSigma^{1/\nu}$, and, for large enough volumes, the expansion can be truncated after the linear term,
\begin{equation}\label{eq:BinderScaling}
    \Binder(\beta_c,m, \NSigma) \simeq  \Binder(\beta_c,\LatMassStaggeredZTwo, \infty) + \textcolor{blue}{c} \, (m - \textcolor{blue}{\LatMassStaggeredZTwo}) \NSigma^{1/\nu}.
\end{equation}
In our case, the critical value for the mass $\LatMassStaggeredZTwo$ corresponds to a second order phase transition in the 3D Ising universality class, so that one can fix $\Binder(\beta_c,\LatMassStaggeredZTwo, \infty) = 1.604$ and $\nu = 0.6301$ to better constrain the fit. 

While, strictly speaking, away from the chiral limit the true scaling field is a mixture of the chiral condensate and gauge variables,
it has been known for a time~\cite{Karsch:2001nf} and demonstrated repeatedly, e.g.~\cite{Jin:2017jjp},
that the scaling of $\Binder(\beta,m)$ is dominated by the chiral condensate,
with corrections due to mixing suppressed by the volume and vanishing in the thermodynamic limit. Sensitivity to mixing can be
checked by comparing the critical exponent with its infinite volume value~\cite{Jin:2017jjp}.
We thus confirmed that fits with $\Binder(\beta_c,\LatMassStaggeredZTwo, \infty)$ and $\nu$ as free parameters give results
consistent with the thermodynamic limit, before fixing them to their known values for precision.

Our simulated values for $\Binder(\beta_c,m, \NSigma)$ are then fitted to Eq.~\eqref{eq:BinderScaling} and the fit parameters $\textcolor{blue}{c}$ and $\textcolor{blue}{\LatMassStaggeredZTwo}$ are extracted.
We are particularly interested in $\LatMassStaggeredZTwo$ indicating the position in mass of the $\ZTwoUniversality$ critical boundary in the $(\LatMassStaggeredZTwo,\Nf)$-plane.
We varied the spatial extent of the lattice $\NSigma$ such that  the aspect ratios, governing the size of the box in physical units at finite temperature, was in the range $L/T=\NSigma/\NTau\in\lbrace2,3,4,5\rbrace$.
The whole study has been repeated for $5$ different $\Nf$ values,  $\Nf\in\lbrace2.8, 2.6, 2.4, 2.2, 2.1\rbrace$.
For each parameter set $\lbrace \Nf, m, \NSigma, \beta \rbrace$, statistics of about $(200k-400k)$ trajectories has been accumulated over $4$ Markov chains, subject to the requirement that the skewness of the chiral condensate distribution is compatible within $2$ to $3$ standard deviations between the different chains.
A detailed overview of all our simulation runs is provided in Table~\ref{tab:simOverview}.

\section{Finite size effects and the zero-crossing of the skewness \label{sec:finiteSizeEffects}}

\begin{figure*}
    \centering
    \subfigure[ $\chiralcond$ histogram at $\Nf=2.8$, $m=0.0200$, $\NSigma=8$ ]
        {\label{fig:hist1}\includegraphics[width=.45\textwidth]{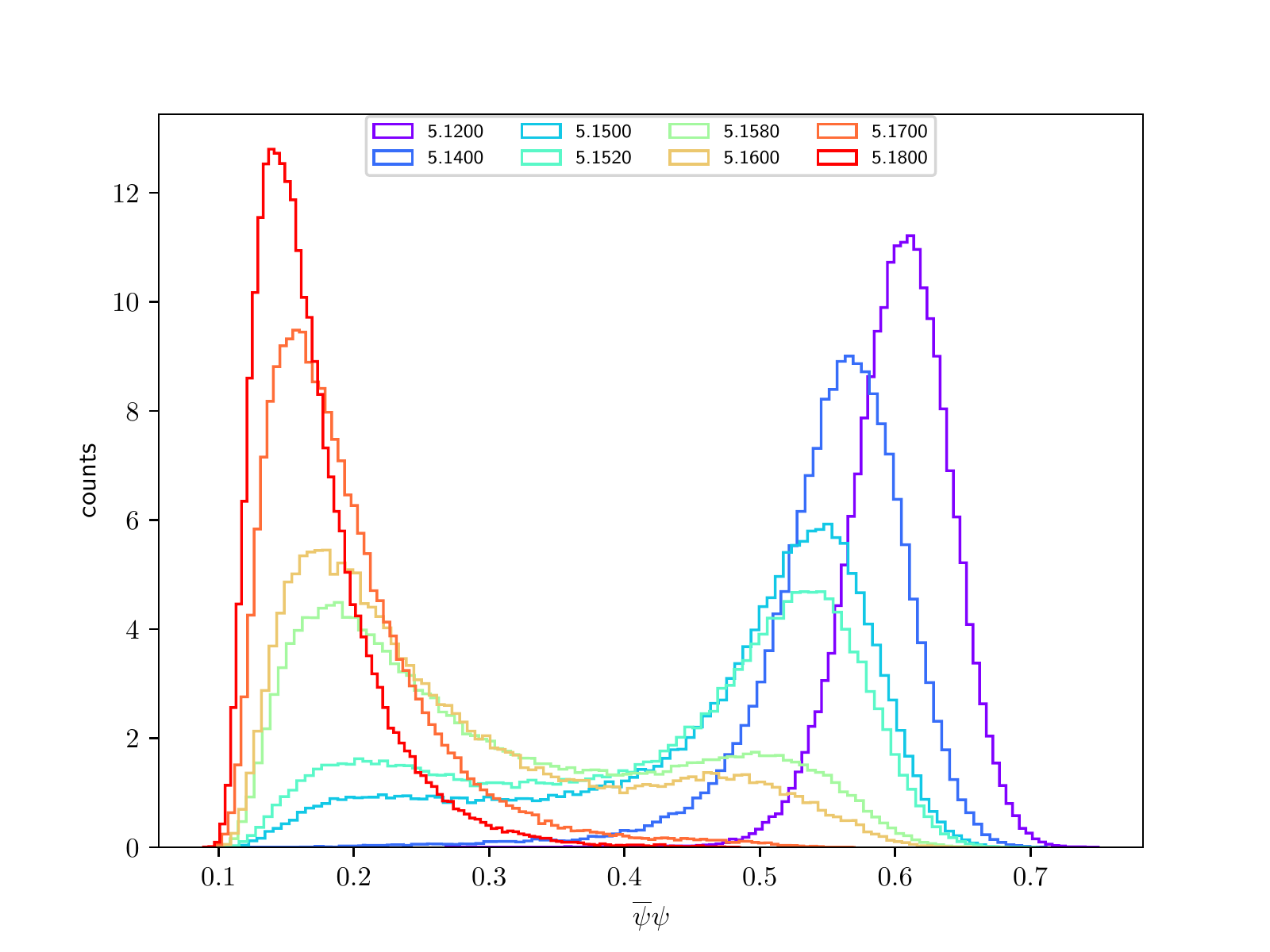}} \qquad
    \subfigure[ $\Skewness(\chiralcond;\beta)$ at $\Nf=2.8$, $m=0.0200$, $\NSigma=8$]
        {\label{fig:skew1}\includegraphics[width=.45\textwidth]{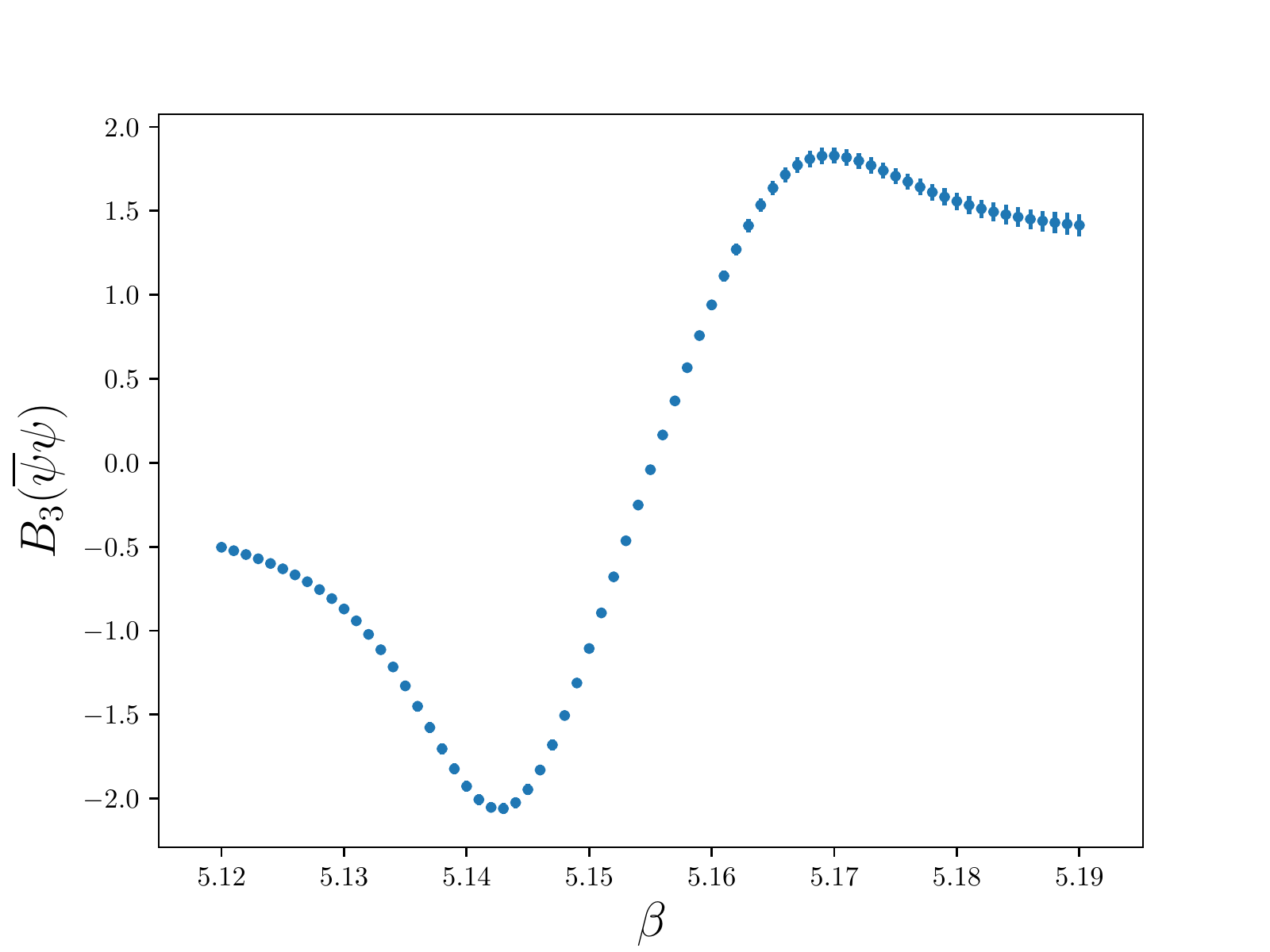}} \\
    \caption{\label{fig:collapseBinderMass0200_8} Examples of chiral condensate normalized histograms and corresponding skewness.}
    \subfigure[ $\chiralcond$ histogram at $\Nf=2.8$, $m=0.0200$, $\NSigma=12$ ]
        {\label{fig:hist2}\includegraphics[width=.45\textwidth]{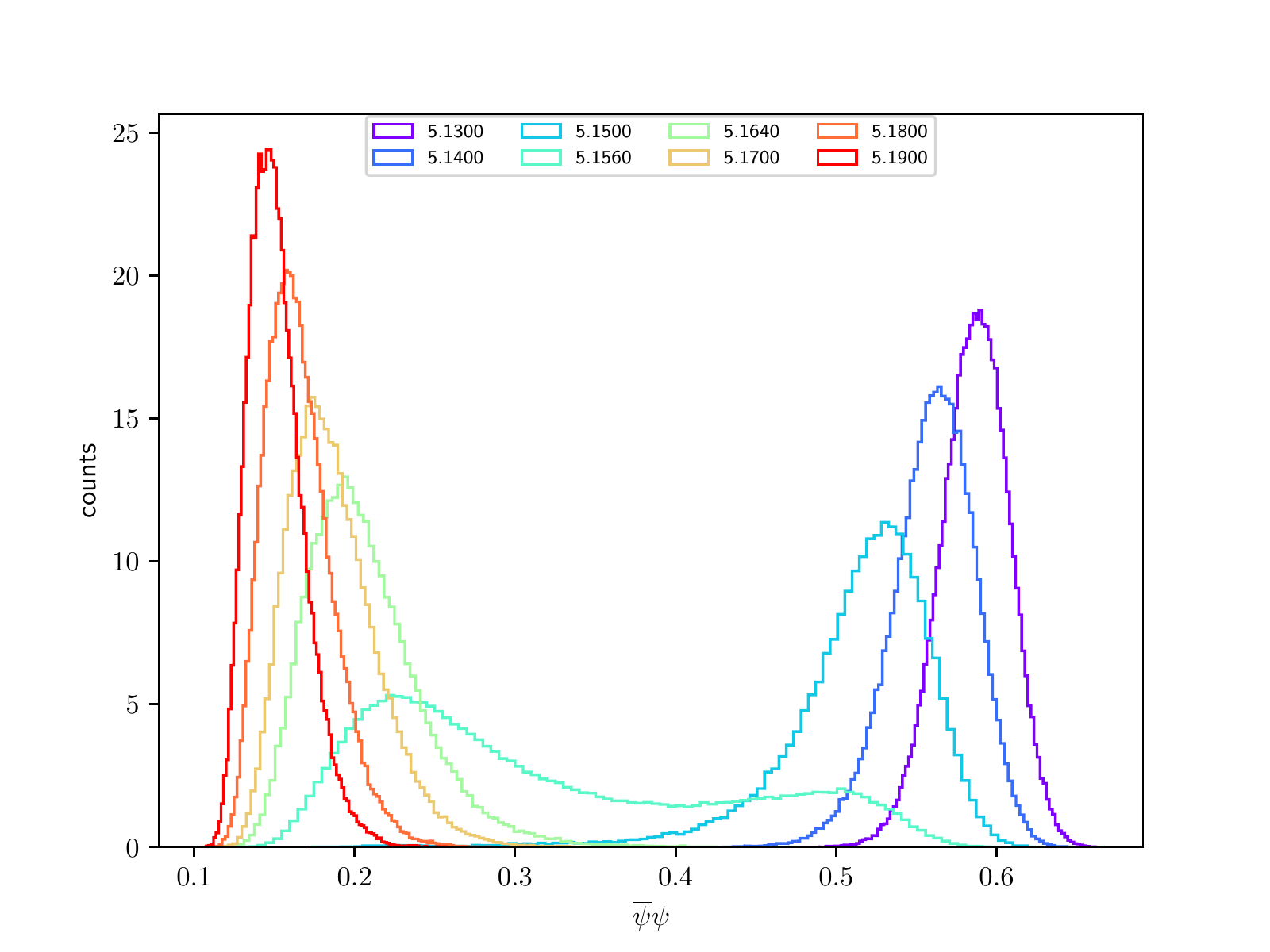}} \qquad
    \subfigure[ $\Skewness(\chiralcond;\beta)$ at $\Nf=2.8$, $m=0.0200$, $\NSigma=12$]
        {\label{fig:skew2}\includegraphics[width=.45\textwidth]{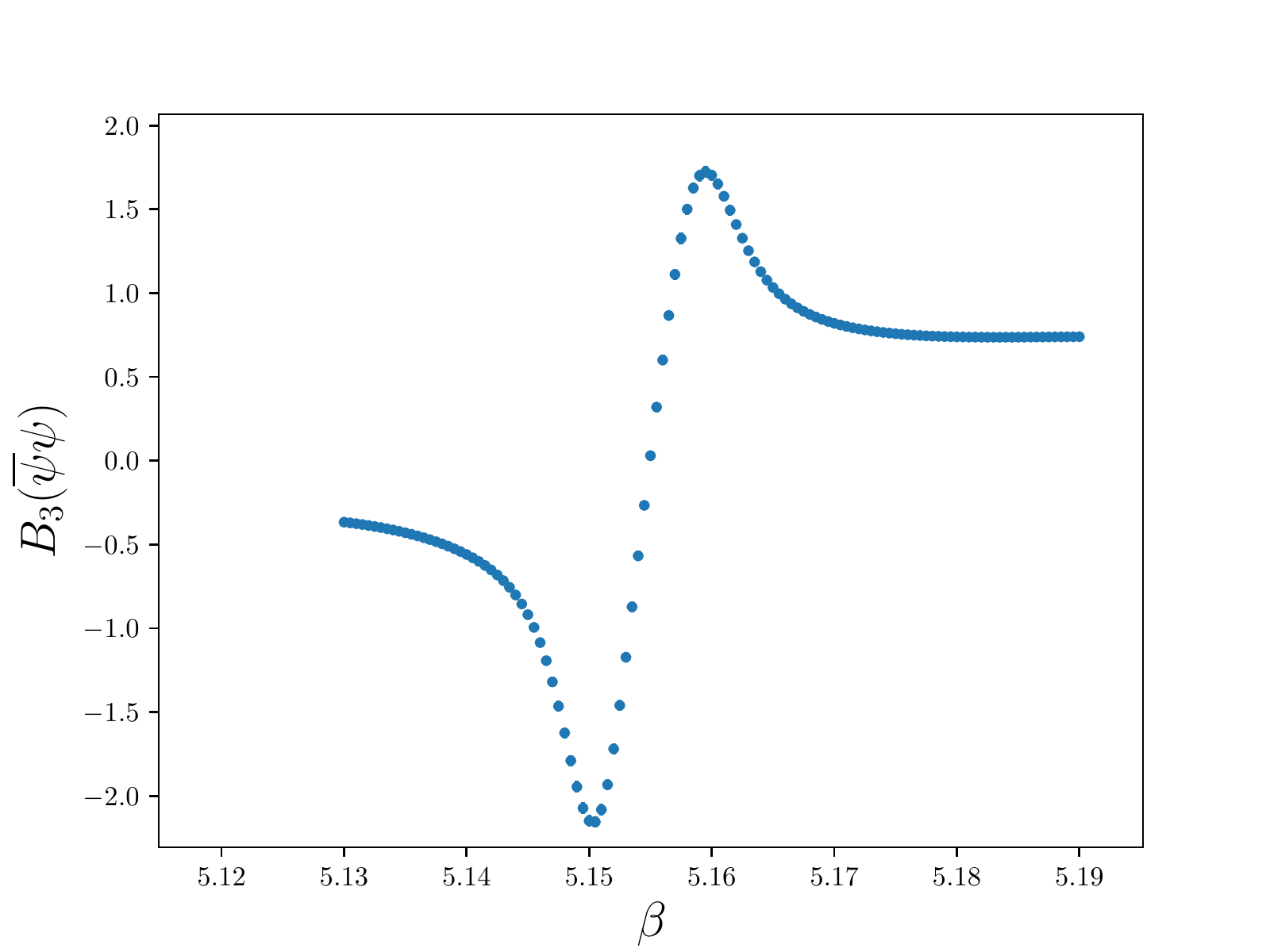}}
    \caption{\label{fig:collapseBinderMass0200_12} Examples of chiral condensate normalized histograms and corresponding skewness.}
\end{figure*}

Besides the increase in simulation time, numerical studies approaching the chiral limit also feature additional intricacies in the data analysis.
We are not aware of any literature on these technical issues and hence present them here.

Below the critical temperature chiral symmetry is both spontaneously and explicitly broken, above the critical temperature there is chiral symmetry restoration for what concerns the spontaneous breaking, still quark masses break it explicitly.
This reflects in our results for the chiral condensate $\chiralcond$ being non-zero both in the chirally broken and in the chirally restored phase.
Before going to actual results let us first consider what happens in some volume $V$ close enough to the thermodynamic limit so that the tunneling probability between broken and restored phase is suppressed.
Then the distribution of measured values for the chiral condensate will always be a single normal distribution of some given width characterized by the mean value $\chiralcond \simeq 0$ ($\chiralcond \ne 0$) for $\beta > \beta_c$ ($\beta < \beta_c$) in the restored (broken) phase.
The third standardized moment of such distributions i.e. its skewness would be zero at every $\beta$.

However, our simulations take place in not so large volumes where tunneling between stable ground states is probable due to the finite height of the potential barrier between them.
This has important effects on the sampled distribution of the chiral condensate and also sets the procedure for identifying the phase boundary.
In formulae one can model the described behavior considering the chiral condensate distribution $\mathcal{P}(x)$ expressed as
\begin{equation}\label{eq:distribution}
    \mathcal{P}(x)\equiv c_1\;\mathcal{N}(\mu_1,\sigma) + c_2\;\mathcal{N}(\mu_2,\sigma)\;,
\end{equation}
where
\begin{equation}
    \mathcal{N}(\mu,\sigma)\equiv\frac{1}{\sigma\sqrt{2\pi}}\;e^{-\frac{(x-\mu)^2}{2\sigma^2}}
\end{equation}
is a Gaussian distribution with mean $\mu$ and variance $\sigma^2$, $\mu_1$ and $\mu_2>\mu_1$ are positive real numbers, as well as $c_1$ and $c_2$, which are the weights of the restored and broken distributions, respectively.
Since, in a finite volume, the phase transition takes place smoothly around $\beta_c$, both the coefficients $c_1$ and $c_2$ will be nonzero and depend on $\beta$, while always fulfilling the condition $c_1+c_2=1$.
To keep the discussion as simple as possible the variance of the two Gaussian distributions is assumed to be the same, which does not spoil the qualitative description of our data.

\begin{figure*}
    \centering
    \subfigure[ $\chiralcond$ histogram at $\Nf=2.2$, $m=0.0050$, $\NSigma=8$ ]
        {\label{fig:hist3}\includegraphics[width=.45\textwidth]{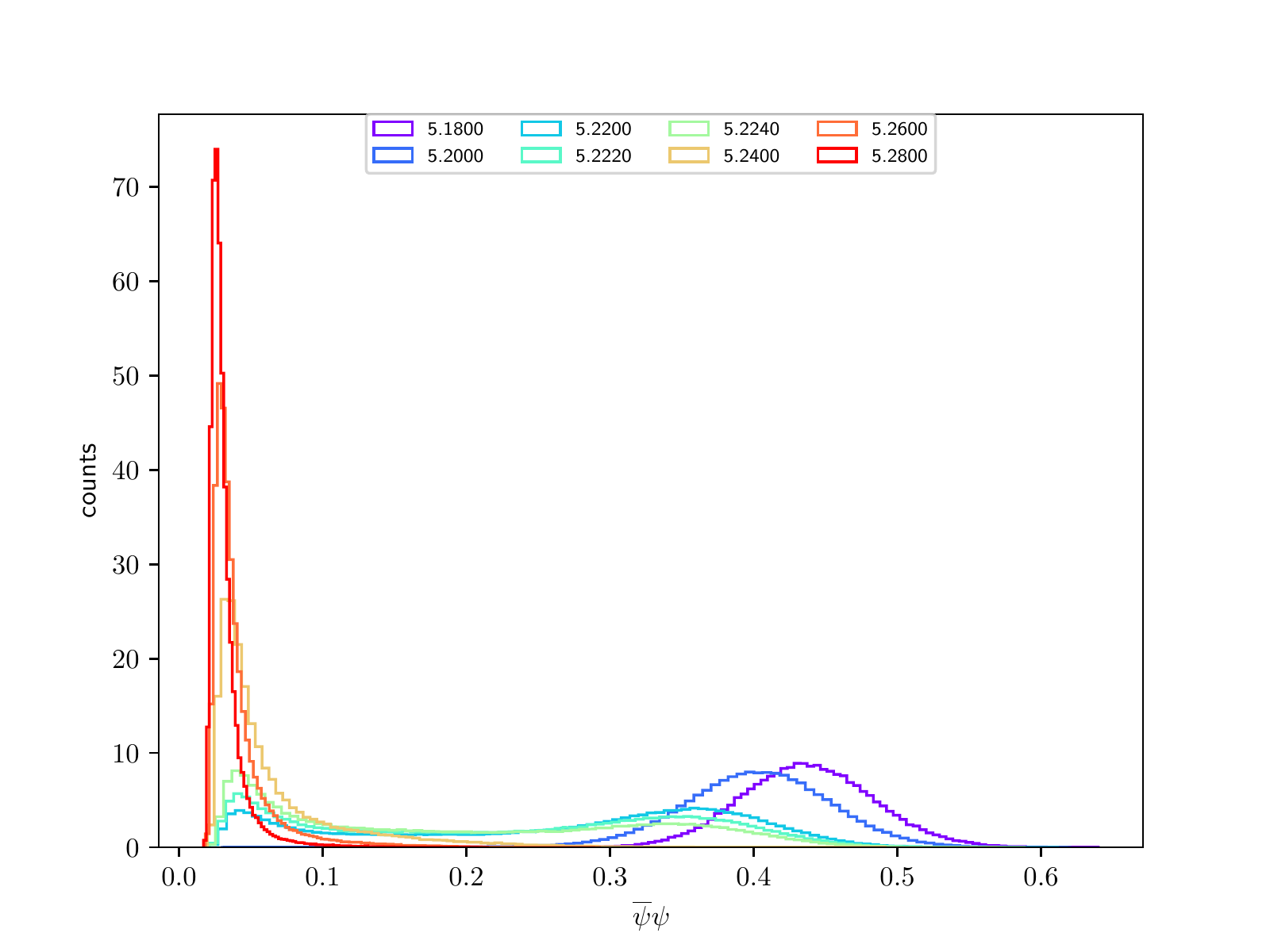}} \qquad
    \subfigure[ $\Skewness(\chiralcond;\beta)$ at $\Nf=2.2$, $m=0.0050$, $\NSigma=8$]
        {\label{fig:skew3}\includegraphics[width=.45\textwidth]{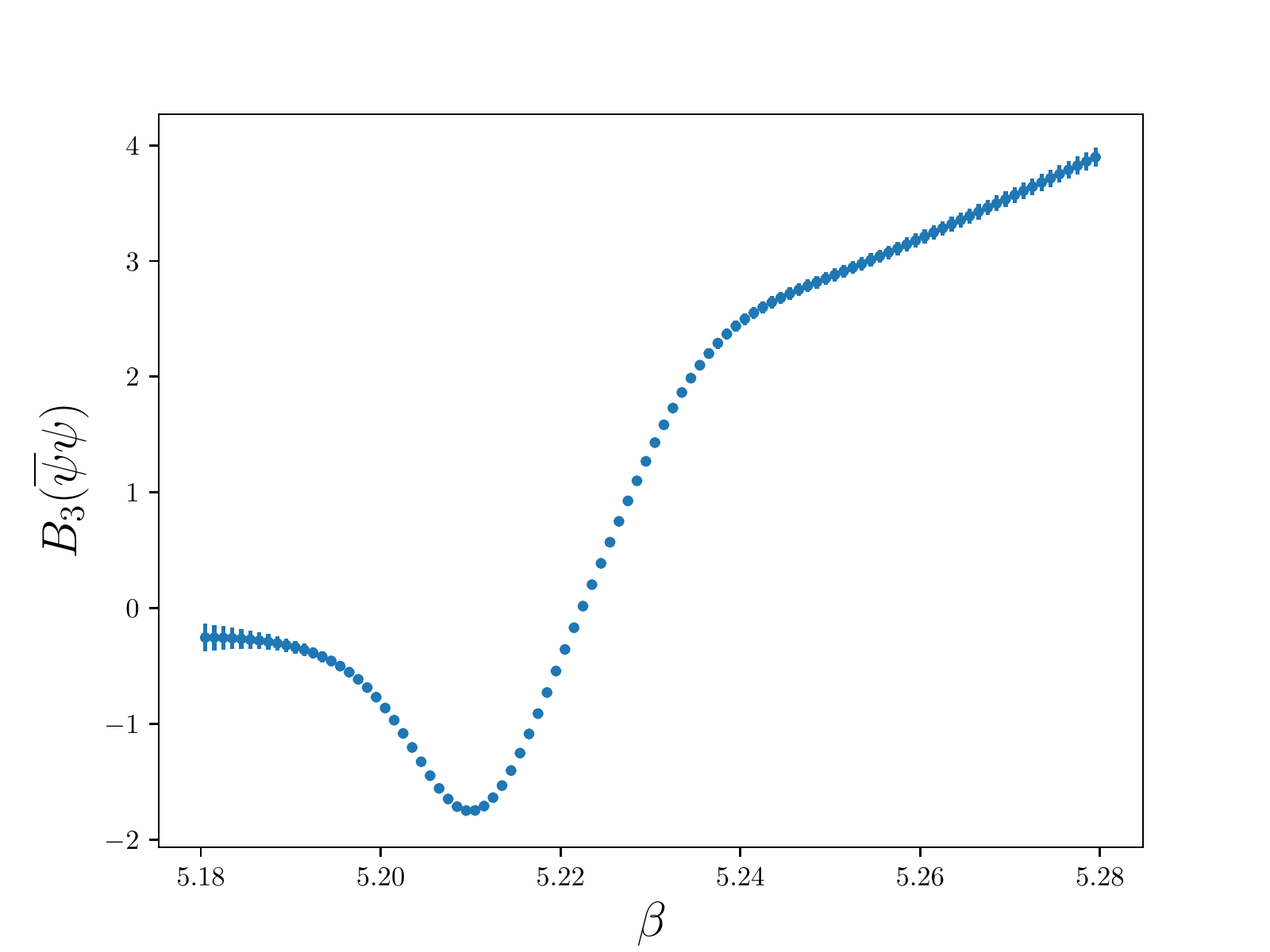}} \\
    \subfigure[ $\chiralcond$ histogram at $\Nf=2.1$, $m=0.0010$, $\NSigma=8$ ]
        {\label{fig:hist4}\includegraphics[width=.45\textwidth]{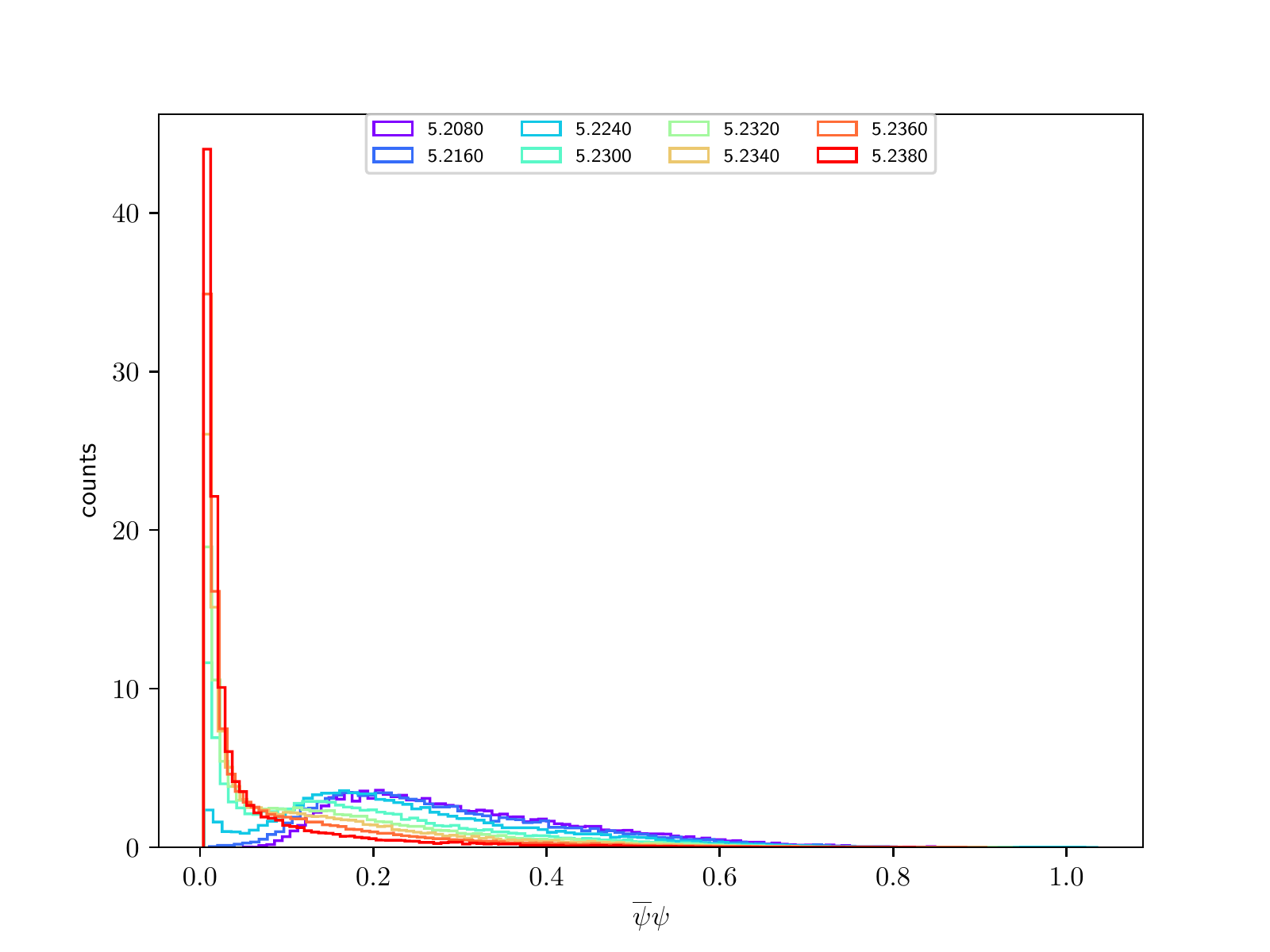}} \qquad
    \subfigure[ $\Skewness(\chiralcond;\beta)$ at $\Nf=2.1$, $m=0.0010$, $\NSigma=8$]
        {\label{fig:skew4}\includegraphics[width=.45\textwidth]{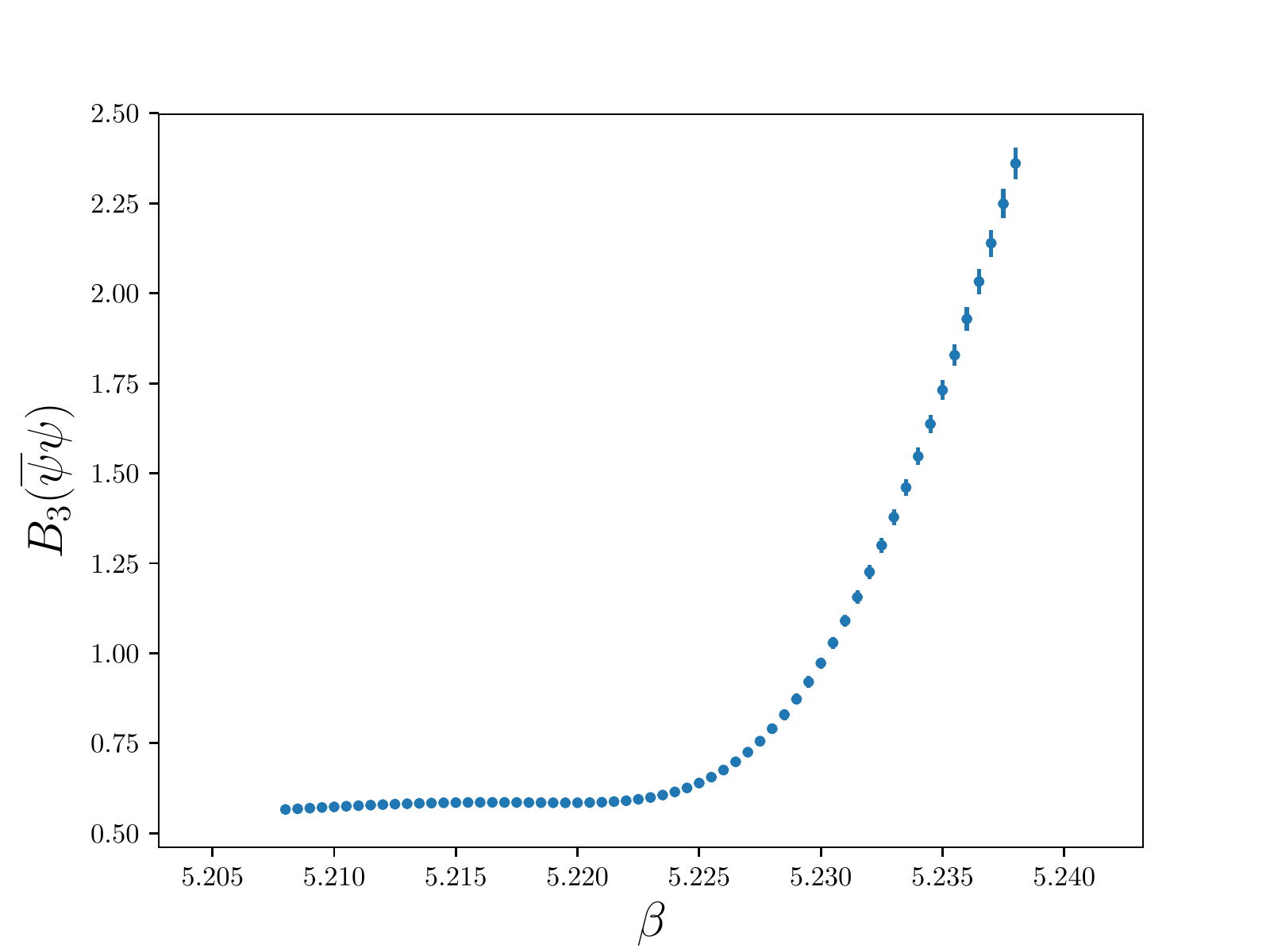}}
    \caption{Examples of chiral condensate normalized histograms and corresponding skewness.}
    \label{fig:collapseBinder}
\end{figure*}

Far away from the critical temperature we will have $c_2=0$ for $\beta \gg \beta_c$ and $c_1=0$ for $\beta \ll \beta_c$.
Approaching the critical temperature, for $\beta \lesssim \beta_c$ we mainly sample in the broken phase (an almost Gaussian peak at $\chiralcond \ne 0$), but also tunnel to the restored phase  (an almost Gaussian peak at $\chiralcond \sim 0$), so that our distribution develops a longer left tail, i.e.~it is left-skewed.
This means that its third moment is negative.
On the other hand, for $\beta \gtrsim \beta_c$ we mainly sample in the restored phase, but still tunnel to the broken phase with some probability.
In that case the sampled distribution is right-skewed, i.e.~its third moment is positive.
Indeed, the general expression for the skewness of $\mathcal{P}(x)$ is
\begin{equation}\label{eq:skewness}
    \Skewness\bigl[\mathcal{P}(x)\bigr]=\frac{\int_{-\infty}^{+\infty} (x-\langle x\rangle)^3\;\mathcal{P}(x)\;dx}
                                                 {\Bigl[\int_{-\infty}^{+\infty} (x-\langle x\rangle)^2\;\mathcal{P}(x)\;dx\Bigr]^\frac{3}{2}}\;
\end{equation}
and using Eq.~\eqref{eq:distribution} leads to
\begin{equation}
   \Skewness\bigl[\mathcal{P}(x)\bigr]=\frac{c_2 (2c_2-1)(c_2-1)(\mu_2-\mu_1)^3}{[\sigma^2+(c_2-c_2^2)(\mu_2-\mu_1)^2]^\frac{3}{2}}\;.
\end{equation}
It is easy to see that $\Skewness\bigl[\mathcal{P}(x)\bigr]<0$ any time $c_2>\frac{1}{2}$ and $\Skewness\bigl[\mathcal{P}(x)\bigr]>0$ for $c_2<\frac{1}{2}$.

For a more detailed picture we also model the $\beta$-dependence of both the coefficient $c_2$ and the peak positions $\mu_1$ and $\mu_2$, imposing the constraints
\begin{align*}
    c_2(\beta\gg\beta_c,\beta_c)=0 \quad&\text{and}\quad c_2(\beta\ll\beta_c,\beta_c)=1\;;\\
    \mu_1(\beta\gg\beta_c,\beta_c)=\overline{x}_1 \quad&\text{and}\quad \mu_2(\beta\ll\beta_c,\beta_c)=\overline{x}_2\;.
\end{align*}
Naive ans\"atze satisfying the above conditions are
\begin{align}\label{eq:ansaetze}
    c_2(\beta,\beta_c) &= \frac{1}{2} \left\lbrace 1-\tanh\left[\xi\left(\beta-\beta_c\right)\right]\right\rbrace\;;\nonumber\\
    \mu_1(\beta,\beta_c) &= \frac{\overline{x}_2-\overline{x}_1}{2} c_2(\beta,\beta_c) + \overline{x}_1\;;\nonumber\\
    \mu_2(\beta,\beta_c) &= \frac{\overline{x}_2-\overline{x}_1}{2} c_2(\beta,\beta_c) + \frac{\overline{x}_1+\overline{x}_2}{2}\;,
\end{align}
where $\xi$ is just a scale factor.
Note, however, that the requirements $\quad c_2(\beta\ll\beta_c,\beta_c)=1$ and \mbox{$\mu_2(\beta\ll\beta_c,\beta_c)=\overline{x}_2$} are only approximately fulfilled for $\beta=0$. Moreover, according to our ans\"atze, we also have
\begin{align*}
    \mu_1(\beta\ll\beta_c,\beta_c)=\frac{\overline{x}_1+\overline{x}_2}{2} \enspace&\text{,}\enspace \mu_2(\beta\gg\beta_c,\beta_c)=\frac{\overline{x}_1+\overline{x}_2}{2}\;.
\end{align*}

The skewness of $\mathcal{P}(x)$ as a function of $\Delta_\beta\equiv(\beta-\beta_c)$ follows by inserting the ans\"atze of Eq.~\eqref{eq:ansaetze} into Eq.~\eqref{eq:distribution} and evaluating Eq.~\eqref{eq:skewness},
\begin{equation*}
    \Skewness\bigl[\mathcal{P}(x)\bigr]= \frac{2(\overline{x}_2-\overline{x}_1)^3 \,\sech\left(\xi\,\Delta_\beta\right)^2 \,\tanh\left(\xi\,\Delta_\beta\right)}{\left[\sigma^2 + (\overline{x}_2-\overline{x}_1)^2 \sech\left(\xi\,\Delta_\beta\right)^2\right]^\frac{3}{2}}\;.
\end{equation*}
Figure~\ref{fig:AnalyticModel} shows the modeled behavior.
The qualitative agreement with our simulation results  for \mbox{$\Nf=2.8$} and \mbox{$m=0.0200$}, Figure~\ref{fig:collapseBinderMass0200_8}, is pretty convincing for \mbox{$\beta \ll \beta_c$} and around \mbox{$\beta_c$}, but not as much for \mbox{$\beta \gg \beta_c$}, where the skewness seems to flatten out at some positive value rather than at zero. Still, we observe our model to give a better and better description of numerical results in a larger and larger volume, see Figure~\ref{fig:collapseBinderMass0200_12}. 

\begin{figure*}[!ht]
    \centering
        {\includegraphics[width=\columnwidth,clip]{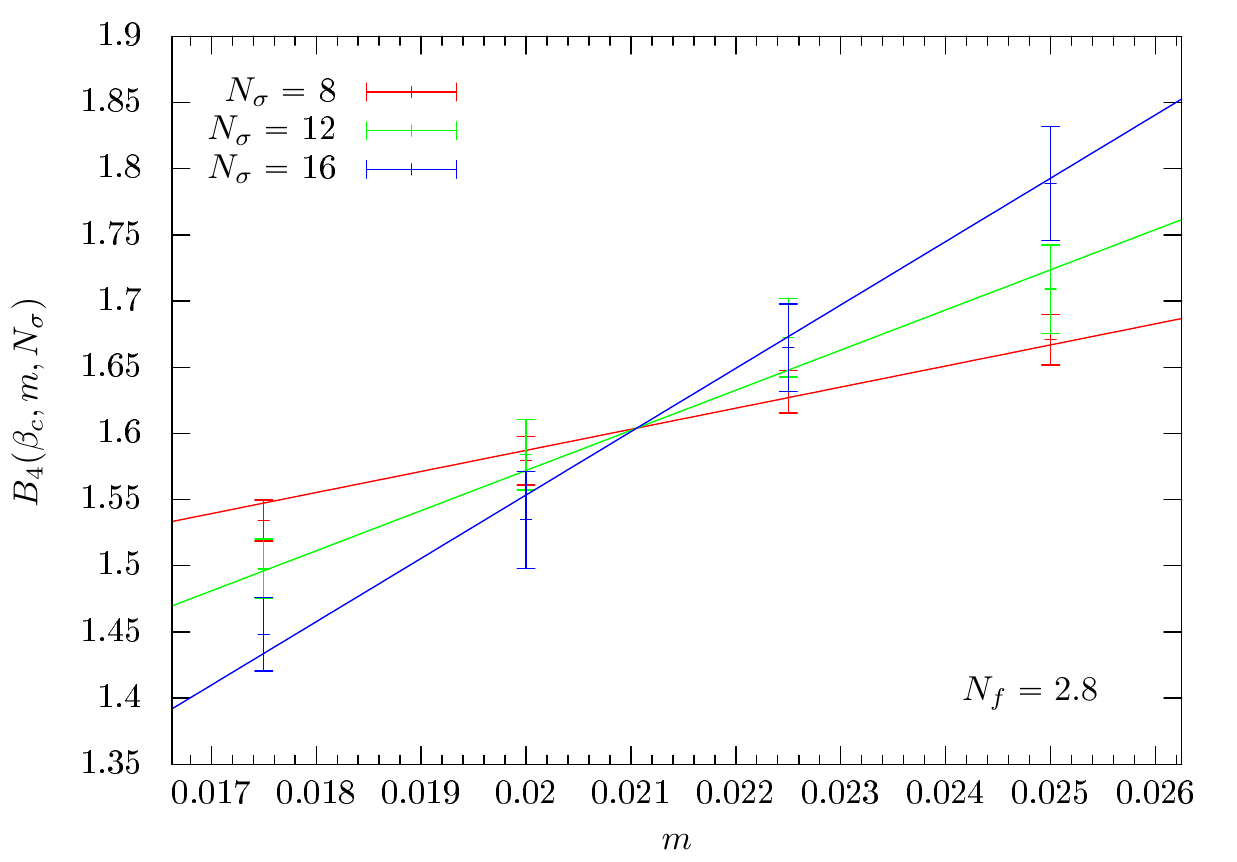}}
        {\includegraphics[width=\columnwidth,clip]{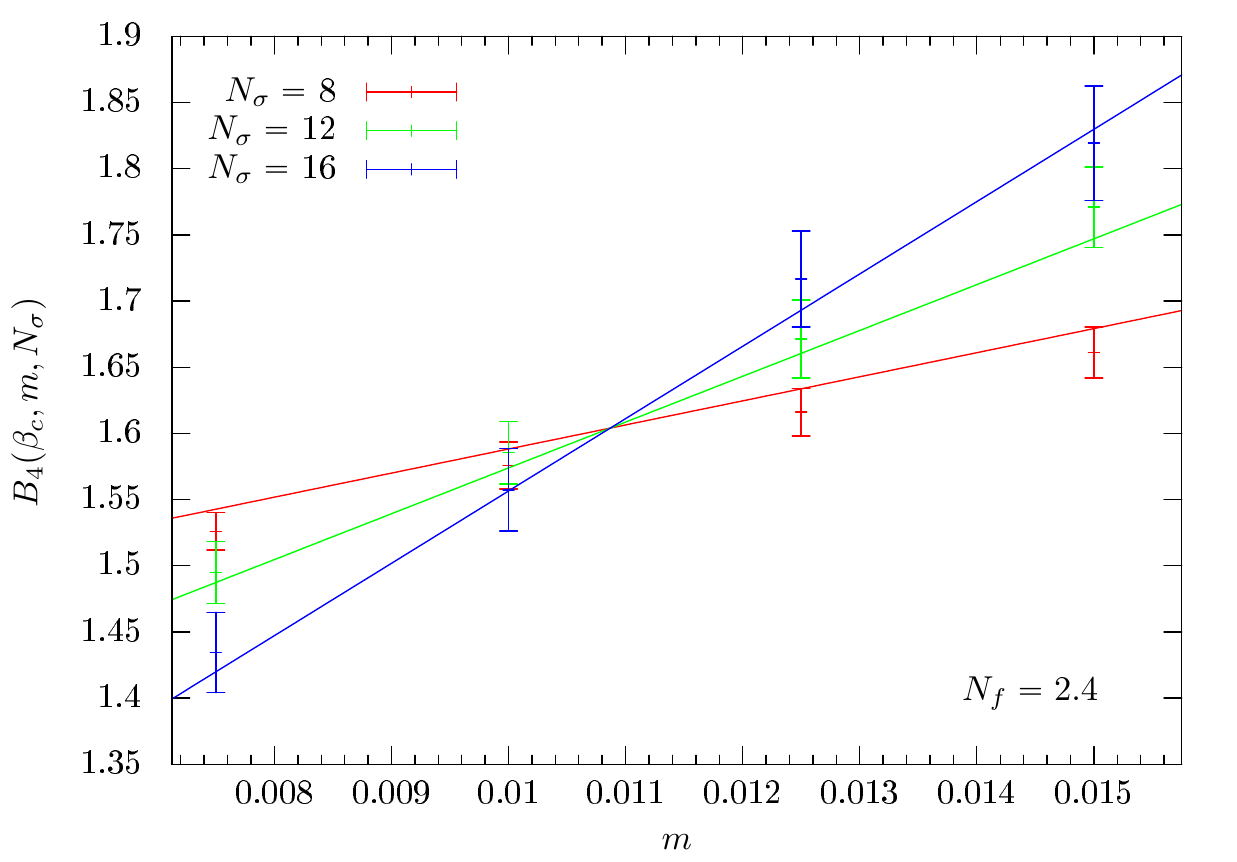}}
    \caption{Finite size scaling of $\Binder$ and fit.\label{fig:binderFit}}
\end{figure*}

On the other hand, what we can observe in Figure~\ref{fig:collapseBinderMass0200_8}, as well as in Figure~\ref{fig:collapseBinder}, is the way this picture gets distorted when the volume gets too small for a decreasing bare quark mass.
The reason for the distortion is the hard lower bound imposed on the distribution of the chiral condensate by our necessarily finite bare quark mass.
What happens is that wider and wider (as the volume gets too small) Gaussian peaks in the restored phase at very small values of the bare quark mass become strongly right-skewed because a symmetric left tail would extend beyond the lower bound.

There can still be a value of the coupling (temperature) at which the overall distribution is symmetric when the peak corresponding to the broken phase is sufficiently populated. However, once only the peak corresponding to the restored phase is sampled, its asymmetry prevents the skewness to become zero again (see Figures~\ref{fig:skew1} and~\ref{fig:skew3}).
At even smaller masses the asymmetry characterizing this peak is such that the overall distribution stays right-tailed, hence the skewness stays positive, at any temperature (see Figures~\ref{fig:hist4} and~\ref{fig:skew4}).
While the volume is increased, both the position of the peaks and their width change in such a way that the expected picture is recovered at any fixed mass.

This illustrates the crucial necessity of having larger and larger volumes while the bare quark mass is decreased, in order to avoid a qualitatively biased analysis.
Chiral perturbation theory also reminds us of the increasing severity of finite size effects at fixed lattice volume while the bare quark mass is reduced.
The GMOR relation
\begin{equation}
    \mpi^2 = \frac{(m_u + m_d)}{F_\pi^2} |\langle0|\overline{u}u|0\rangle|,
\end{equation}
tells us that, in order to maintain $\mpi L \gg 1$, $L$ should grow with the square root of the factor by which the sum of the quark masses is decreased.

This observation on finite size effects explains the huge growth in the cost of identifying the critical temperature at smaller and smaller bare quark mass (i.e.~ smaller $\Nf$).
These considerations also have to be taken into account in estimating the costs of extending the present study to larger $\NTau$.
In our investigation we used the smallest volume at which a zero crossing of the skewness could be found as the minimal volume in our finite-size scaling analysis. However, to check that the residual finite size effects ($\Skewness \ne 0$ for $\beta \gg \beta_c$) were not affecting our results
for the critical mass in the case of $\Nf=2.1$, we performed simulations over $4$ different volumes (see Table~\ref{tab:simOverview}), considering also $\NSigma=24$ and then excluding in turn the biggest and the smallest volume from the fit. In this way we could check the stability in the value of  $\LatMassStaggeredZTwo$.

\section{Results \label{sec:results}}
Our results for $\LatMassStaggeredZTwo$ for all considered $\Nf$ are collected in Table~\ref{tab:fitResults} and they are the outcome of the kurtosis fit procedure discussed in Section~\ref{sec:num}, of which examples, for the cases $\Nf=2.8,2.4$, are plotted in Figure~\ref{fig:binderFit}.
The ordering of kurtosis values for a fixed mass $m$ as function of the volume depends, as expected, on whether $m$ lies in the crossover region ($\Binder$ increases with $\NSigma$) or in the first order region ($\Binder$ decreases with $\NSigma$), $\LatMassStaggeredZTwo$ being the mass at which sets of $\Binder$ values for different volumes cross in one point (provided we are effectively in the thermodynamic limit).

\begin{figure*}
    \centering
    \subfigure[\label{fig:nonRescaled} $\ZTwoUniversality$ critical line in the $(m,\Nf)$-plane ]
        {\includegraphics[width=0.985\columnwidth,clip]{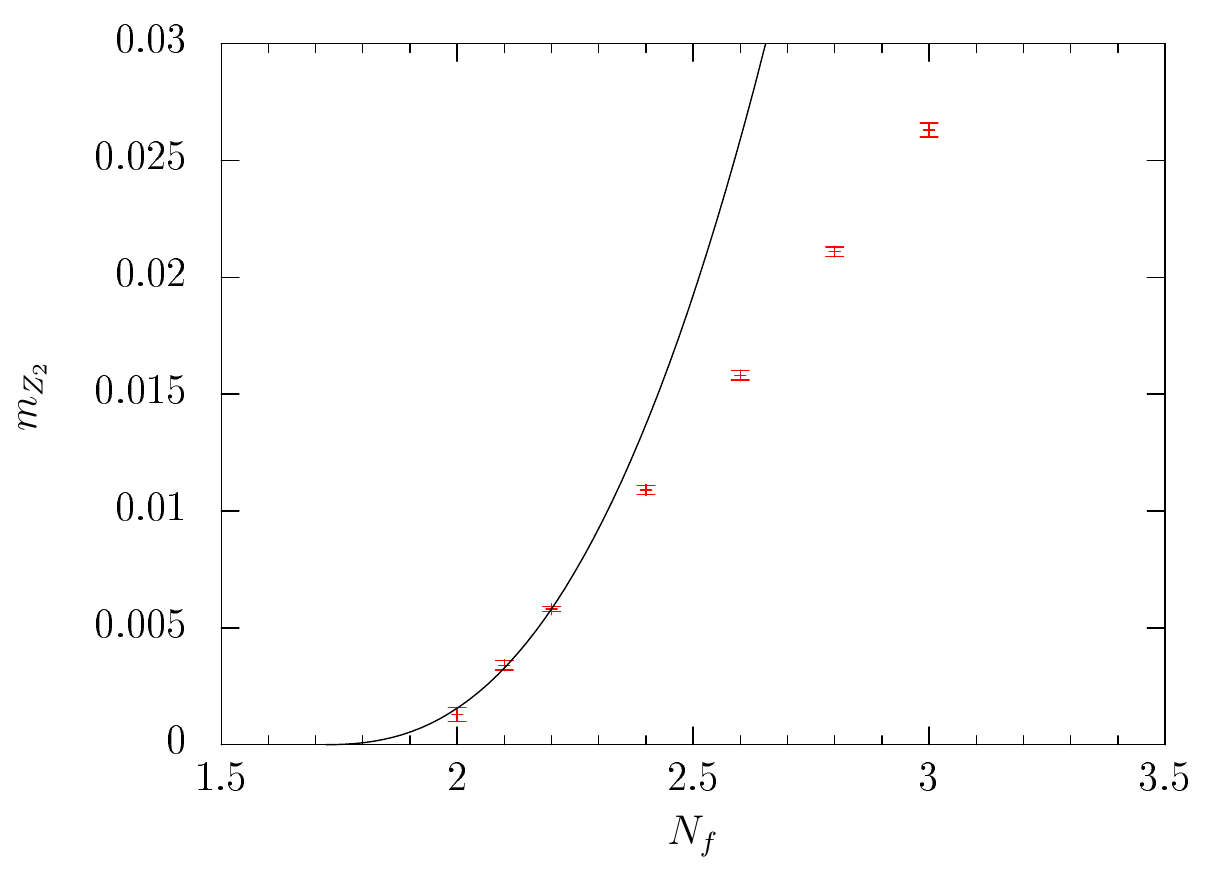}}
    \hfill
    \subfigure[\label{fig:rescaled} $\ZTwoUniversality$ critical line in the $(m,\Nf)$-plane ]
        {\includegraphics[width=\columnwidth,clip]{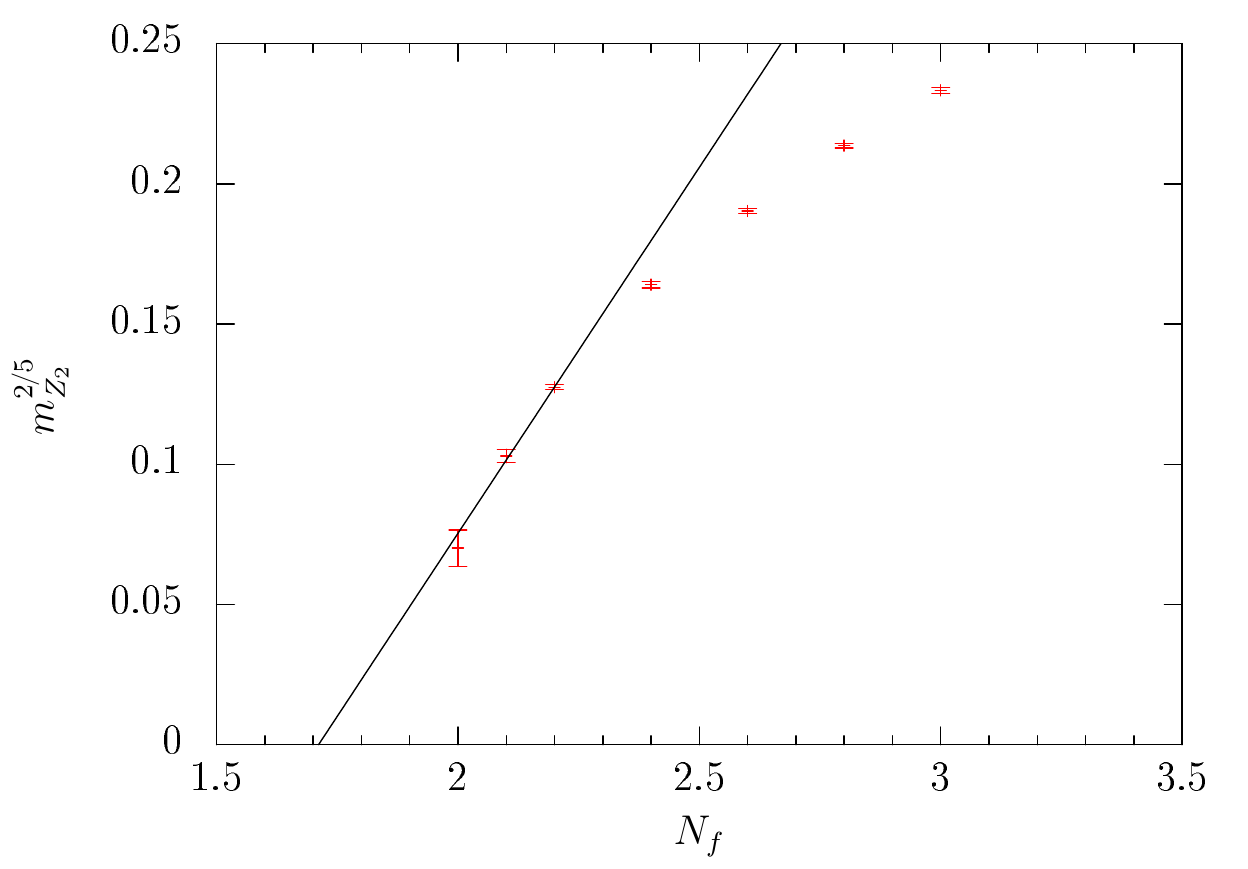}}
    \caption{A linear fit in the range $[2.0, 2.2]$ is displayed in the rescaled plot (\subref{fig:rescaled}) and the corresponding curve is drawn in the non-rescaled plot (\subref{fig:nonRescaled}).
    For $\Nf=2.0$ we use the result from the extrapolation from imaginary chemical potential~\protect\cite{Bonati:2014kpa}.\label{fig:tricrScaling}}
\end{figure*}

\renewcommand{\arraystretch}{1.2}
\begin{table}[!ht]\centering
    \begin{tabular}{cccccc}
        \toprule[0.3mm]
        $\Nf$ & $a m$ - range & $\LatMassStaggeredZTwo$ & $c$  & $\chi^2$  \\
        \midrule[0.1mm] 
        3.0  & [0.0250:0.0450] & 0.0259(9) & 0.51(6) & 0.19 \\
        2.8  & [0.0175:0.0250] & 0.0211(3) & 0.59(5) & 0.27 \\
        2.6  & [0.0125:0.0200] & 0.0158(2) & 0.60(5) & 0.18 \\
        2.4  & [0.0075:0.0150] & 0.0109(2) & 0.67(5) & 0.55 \\
        2.2  & [0.0025:0.0100] & 0.0058(1) & 0.77(4) & 0.27 \\
        2.1  & [0.0015:0.0045] & 0.0034(2) & 0.88(14)& 0.15 \\
        \bottomrule[0.3mm]
    \end{tabular}
    \caption{Results for fits to Eq.~\protect\eqref{eq:BinderScaling}. The $\Nf=3.0$ result is taken from~\cite{deForcrand:2006pv}. $\Binder(\LatMassStaggeredZTwo,\infty)$ and $\nu$ have been set to $1.604$ and $0.6301$, respectively.}
    \label{tab:fitResults}
\end{table}

Table~\ref{tab:fitResults} and Figure~\ref{fig:tricrScaling} show $\LatMassStaggeredZTwo$ to decrease continuously towards zero as $\Nf$ is lowered.
We observe a strinking near-linear behavior down to very small quark masses.
Regarding the possibility of future continuum extrapolations, it would be very interesting to find an explanation for this behavior and whether it is possibly genuine on finer lattices as well.

As the critical line approaches the chiral limit we employ the fact  that, in the vicinity of a tricritical point, it has to display a power law dependence with known critical exponents~\cite{lawrie}.
The scaling law in our case is
\begin{equation}
    m^{2/5}_{\ZTwoUniversality}(\Nf) = \textcolor{blue}{C}\left(\Nf - \textcolor{blue}{\NfTricr}\right). 
\end{equation}
The plot 
of the rescaled mass $m^{2/5}_{\ZTwoUniversality}$ as a function of $\Nf$ is displayed in Figure~\ref{fig:rescaled}.
Our data are consistent with the expected scaling relation for $\Nf\le2.2$ and hence provide an extraplation to the chiral limit resulting in $\NfTricr<2$.

Finally, note that our simulated results for \mbox{$\Nf=2.2,2.1$} are fully aligned with the result for $\Nf=2.0$, which is based on a similar tricritical extrapolation from imaginary chemical potential~\cite{Bonati:2014kpa}.
This is an independent validation of the former extrapolation and at the same time confirms this point to be within the tricritical scaling window, 
 whose width in mass appears to be roughly the same in both extrapolations. 
Moreover, we even attempted a direct extraction of the critical mass $\LatMassStaggeredZTwo$ at $\Nf=2.0$ to cross-check the result obtained in the extrapolation from imaginary chemical potential (see Table~\ref{tab:simOverview}). The increase in simulation time was so high that we could not complete the extraction of $\LatMassStaggeredZTwo$ via fit of the kurtosis. On the other hand, for all masses and volumes considered the measured value for  $\Binder(\beta_c,m,\NSigma)$ turned out to be compatible within errors with $\Binder(\LatMassStaggeredZTwo,\infty)$.
This further illustrates the benefit of the tricritical extrapolation discussed here.

\section{Conclusions \label{sec:conc}}
On general symmetry grounds, the chiral phase transition strengthens with the number of light quark flavors.
On coarse lattices with unimproved actions it is known to be of first-order for $\Nf\geq 2$.
Here we investigated the nature of the chiral phase transition in the $(m,\Nf)$-plane with $\Nf$ interpreted as a continuous parameter in the path integral formulation of the theory.
If the transition in the chiral limit changes from first order to second by reducing $\Nf$, there has to exist a tricritical point at some $\NfTricr$.
The extrapolation of the second order boundary line between the first-order and crossover regions to the chiral limit is then constrained by universality.
Our data are indeed consistent with tricritical scaling in the range $\Nf\in[2.0,2.2]$ and offer a way to extrapolate to the chiral limit in a controlled way.
On our coarse lattices the conclusion is for $\Nf=2$ to be in the first-order region, in agreement with earlier results using imaginary chemical potential instead of $\Nf$ in the same strategy \cite{Bonati:2014kpa}.
Taken together, these might provide useful tools on finer lattices, where ever smaller critical masses need to be identified on the way to a continuum extrapolation.

\begin{acknowledgments}
We thank Philippe de Forcrand, Gergely Endr\"odi and Wolfgang Unger for useful discussions.
The project received initial support by the German BMBF under contract no. 05P1RFCA1/05P2015 (BMBF-FSP 202).
The authors acknowledge support by the Deutsche Forschungsgemeinschaft (DFG) through the grant CRC-TR 211 ``Strong-interaction matter
under extreme conditions'' and by the Helmholtz International Center for FAIR within the LOEWE program of the State of Hesse.
We also thank the staff of \Loewe\ and \Lcsc\ for their support.

\end{acknowledgments}

\bibliographystyle{apsrev4-1}
\bibliography{bibliography}

\onecolumngrid

\end{document}